# Best of both worlds: Synergistically derived material properties via additive manufacturing of nanocomposites

*Mia Carrola[1,2], Amir Asadi,[1,2], Han Zhang[3], Dimitrios G. Papageorgiou[3], Emiliano Bilotti[3], Hilmar Koerner[4]*

[1] Manufacturing and Mechanical Engineering Technology-Department of Engineering Technology and Industrial Distribution, Texas A&M University, College Station, TX 77843-3367, USA

[2] Department of Materials Science and Engineering, Texas A&M University, College Station, TX 77843-3367, USA

[3]School of Engineering and Materials Science, Queen Mary University of London, Mile End Road, London E1 4NS, UK

[4]Materials & Manufacturing Directorate, Air Force Research Laboratory, WPAFB, Ohio 45430, USA

**Abstract:**

With an exponential rise in the popularity and availability of additive manufacturing (AM), a large focus has been directed towards research in this topic's movement, while trying to distinguish themselves from similar works by simply adding nanomaterials to their process. Though nanomaterials can add impressive properties to nanocomposites (NCs), there are expansive amounts of opportunities that are left unexplored by simply combining AM with NCs without discovering synergistic effects and novel emerging material properties that are not possible by each of these alone. Cooperative, evolving properties of NCs in AM can be investigated at the processing, morphological, and architectural levels. Each of these categories are studied as a function of the amplifying relationship between nanomaterials and AM, with each showing the systematically selected material and method to advance the material performance, explore emergent properties, as well as improve the AM process itself. Innovative, advanced materials are key to faster development cycles in disruptive technologies for bioengineering, defense, and transportation sectors. This is only possible by focusing on synergism and amplification within additive manufacturing of nanocomposites.

## Contents

# 1 Background:

Additive Manufacturing, originally described as solid freeform fabrication and rapid prototyping, was initially developed in the late 1980's and early 1990's and uses innovative technologies, such as CAD, CAM, robotics, automation, high rate, and precision to improve processes and products[1]. It plays a critical role in areas where traditional manufacturing is labor intensive and costly due to expensive tool requirements. Recent reviews and news articles claim that additive manufacturing will grow and compete with traditional manufacturing processes.[2] However, this should not be the role of additive manufacturing. While the process will mostly fall short in speed and manufacturing of high volume and volume rates, it does provide solutions as a complementary process that allows manufacturing of parts and components that are impossible to produce with conventional methods. Overall cost in low volume manufacturing might lead to a replacement of conventional processes mostly dictated by the high costs of complexity.

Additive manufacturing has often been termed as 3D printing, although this description is misleading because processes are in fact 2.5D or layer-by-layer assemblies. While 3D bodies are manufactured, the properties are anisotropic due to this limitation. New engineering solutions are emerging in which multi-axis print head technologies are implemented to address the 2.5D limitation by using robotics to get to true 3D deposition of materials.[3-5]

In addition, the line between conventional layer-by-layer processes and AM is becoming more diffuse, especially were processes such as automated fiber or tape placement (AFP, ATP) are miniaturized and merged with AM engineering solutions [6]. Unlike AFP though, additive manufacturing processes described under the ASTM categorization do not require tooling in the traditional sense (support material is still required depending on complexity of the part). Other AM processes such as Composites Based Additive Manufacturing (CBAM, Impossible Objects), follow the original laminated object manufacturing (LOM) processes. [7] Pre-treated sheets of composite materials are compression fused into 3D objects by blasting away untreated material in a final step.

The ability of AM to produce previously un-manufacturable designs combined with compelling economic and lead time benefits has led to significant interest in industry and government to invest in many areas, including soft materials (flexible electronics, tissue engineering, prosthetics) and composites (polymer matrix composites, ceramics matrix composites, metals). Additive manufacturing is a vastly growing field with an exponentially increasing number of publications (**Figure 1**).[8-22]

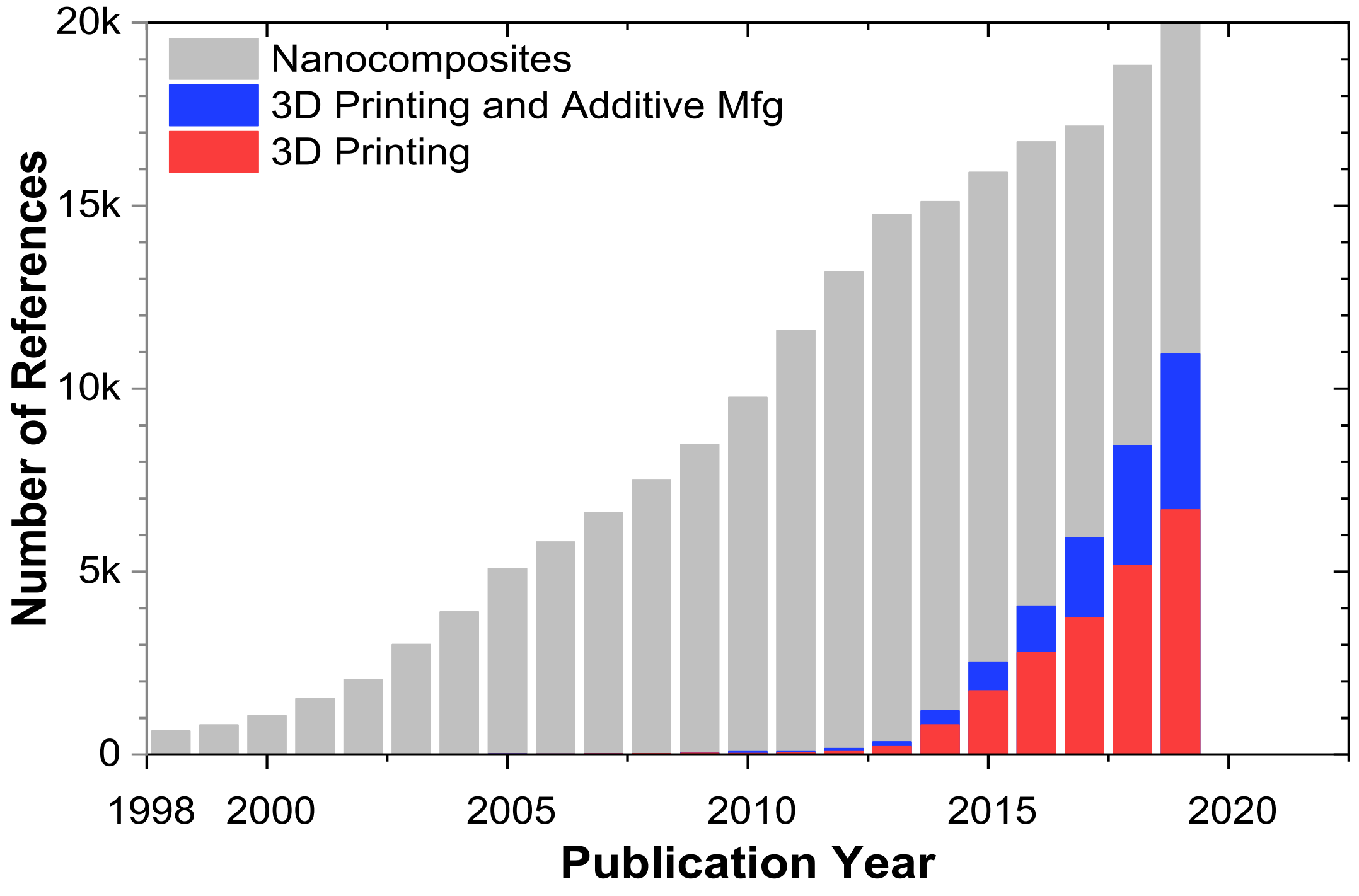

**Figure 1**. Count of publications of the last 22 years for keywords nanocomposites (gray), 3D printing (red) and 3D printing and additive manufacturing combined (blue).

Compared to review articles from recent years with focus on nanocomposites in additive manufacturing, this review will primarily look at research implementing nano-elements into additive manufactured parts either to control processing or to obtain new functions of manufactured parts that are not possible via conventional processing routes.[2, 23-31] The emphasis here is on synergistic effects of nano plus additive manufacturing and not exploring well-established phenomena on nanocomposite properties merely in new confinement geometries, which are dictated by deposition methods. Furthermore, the review article will focus on larger printed parts and only briefly touch on additive manufacturing methods on the micron, sub-micron scale or even molecular printing.[32] In addition, the fast, non-equilibrium processing space necessitates implementation of novel in situ metrology controls, which can be realized by employing nano-elements as probes. The importance of in situ monitoring is summarized in a recent NIST roadmap for polymer-based additive manufacturing. The report offers detailed analyses of the complexities surrounding material characterization, process modeling, in situ measurement, performance, and other cross-cutting challenges for polymer-based AM.[33] It offers a future vision for polymer-based additive manufacturing that implements a digital thread starting from process modeling and is carried into material characterization and in situ measurement to performance of printed parts. In particular, it discussed the development of models to describe and predict the effect of additives to the AM process is listed as a long-term (> 5 years) goal.

We believe that reaching the full potential of additive manufacturing is still a long way ahead. Current problems and issues include reproducibility, the combination of speed and size, resolution, and poor mechanical properties. Recently conducted round robin tests on fused deposition modeled (FDM®) thermoplastic materials highlight the issue of reproducibility with vastly different results from the same equipment conducted in different laboratories, between two pieces of the same equipment within a single laboratory and even between builds in different build plate locations of the same equipment.[34, 35] New material, equipment solutions, and refinements are currently being explored to address these shortcomings. Although the deposition step in additive manufacturing is slow compared to, e.g. injection molding, additive manufacturing allows design complexity that is impossible to obtain through injection molding or similar processes. The main concern is that speed may prohibit large scale production of additively manufactured parts, which may be addressed via modular, interlocking designs of smaller parts.[1, 36] With fused deposition modeled (FDM®) process as an example, there are immediate issues with the layer-by-layer process which leads to sufficient bonding of roads within layers, but also weak bonding between layers (typically z-direction). Each resulting porosity and weak road-to-road bonding may be addressed by incorporation of nanofillers that aid both during processing and post-processing, while also enabling other desired properties, such as external field (electrical, magnetic, acoustic) responsive properties.

These general problems within the deposition based additive manufacturing processes also apply to selective laser sintering (SLS) and stereolithography (SLA). The opportunity space for nano-elements to improve the additive manufacturing process is multi-fold and includes the ability to modify rheology, thermal conductivity and energy transfer, speed-up of post-processing, reduced shrinkage/heat distortion, delamination, and the ability to act as nano-probes for in situ monitoring (dimensional, temperature, strain, alignment).

While it is more obvious to imagine how nanoparticles could aid in the additive manufacturing process, it may also be useful to look at how additive manufacturing offers opportunities to advance nanocomposites. The advancement of nanocomposites requires a synergistic effect, which may include voxelated morphologies enabled by nano-elements that allow discrete, localized property manipulation to address certain functions of the printed part (fold, reinforcement, RF response, to name a few).[37] In general, the ability to build complex hierarchical structures and assemblies with nano-elements that are otherwise not obtainable through conventional processing leads to true advances in nanocomposites.

A combination of both nanocomposite enabled additive manufacturing and additive manufacturing enabled nanocomposites is achievable with strategic choice of nanoelements, e.g. a nanoelement that enables thixotropic properties during processing but also anisotropic mechanical reinforcement in the printed part. A nanoparticle that only acts as rheology modifier might be cause for degradation of mechanical and thermal properties after printed parts are finished.

A common denominator is to control hierarchical morphology and utilize the print path topology to create architectures that are not attainable via conventional processing. Complex, intricate structures already exist in nature. It is therefore not surprising that bioengineering and the medical research field quickly picked up 3D printing techniques to create parts that were impossible to manufacture before (prosthetics, bone structures, soft tissue). On the other hand, the introduction of additive manufacturing into well-established processes and manufacturing requires a paradigm shift in the design process that leads to the desired parts and components. Topology and design optimization for a targeted function (damping, weight, reinforcement, and CTE) is therefore key to truly utilize additive manufacturing for nanocomposites or any other material suite and constitutes a barrier for AM to enter other application areas.[38] A simple search for 3D printing in combination with nano reveals that most publications therefore come from the bioengineering and biochemistry community.

In general, within the constraints of the additive process (i.e., speed, deposition rate, heating/cooling rates, etc.), applied forces during processing are considered to control hierarchical morphology via a counterbalance to viscoelastic forces (gelation & vitrification) to control kinetics of phase separation, particle segregation, particle mobility, chain alignment/crystallization, particle orientation, particle clustering, etc. These processing forces, in the context of nanoparticle filled polymers and resins, are used to control site-specific phenomena such as:

- Matrix phase separation kinetics (e.g., polymer blend in which NP-philic phase or interface is used to direct NP concentration / morphology)
- NP dispersion stability and agglomeration kinetics
- Matrix & particle crystallization
- Local orientation kinetics of particles and polymer chains
- Physical aging of polymer within rigid nano-inclusions
- Mitigating impact of NP at interfaces on chain entanglement across interface; surface elimination, elimination of voids
- Controlling temporal aspects of highly non-linear rheology of highly filled systems (shear thinning rates, shear thickening rates, magnitudes of yield stress, etc.)
- Road-to-road and layer-to-layer entanglements and crosslinking (remotely triggered heating to weld interfaces, RF, H)

**Figure 2** illustrates the overall complexity, but also emphasizes emerging unique opportunities of additive manufacturing with nanoelement inclusions. Under ASTM F2792, five main categories were listed that apply to polymer-based additive processes, namely: a) binder jetting, in which a solution is jetted onto a powder to bind in selected areas; b) extrusion, in which material is deposited through a nozzle or orifice; c) material jetting, in which an aerosol of material solutions is deposited (much like inkjet printing); d) powder bed fusion, in which powder particulates are melt-bonded by heat; and e) photopolymerization, in which a liquid photo-curable resin is selectively cured.[39] We illustrate the interplay of AM with nanoparticles using extrusion-based processes. Other AM processes here have been discussed extensively in other review articles. Three areas are highlighted in which nanoelements can prescribe properties of a printed material to gain optimal performance (flow, mechanical). This includes interplay of forces within morphology during AM of NCs with three sub areas of morphology control that emerge from this simplistic view of the process:

- directed via the additive process (rheology, shear)
- directed via in-line external fields (E, H, UV)
- directed via post-processing (annealing, crosslinking, aging)

Morphology control at these levels is only one aspect of synergistically applying additive manufacturing process using nanocomposites. The main body of the article will review the literature in terms of process control, morphology control and architecture control, three levels in which nanocomposites play an important role in advancing the performance beyond conventionally manufactured parts and components. While this will help with structuring the review, there is work that cannot strictly be binned into one of these categories and may be discussed in more than one level.

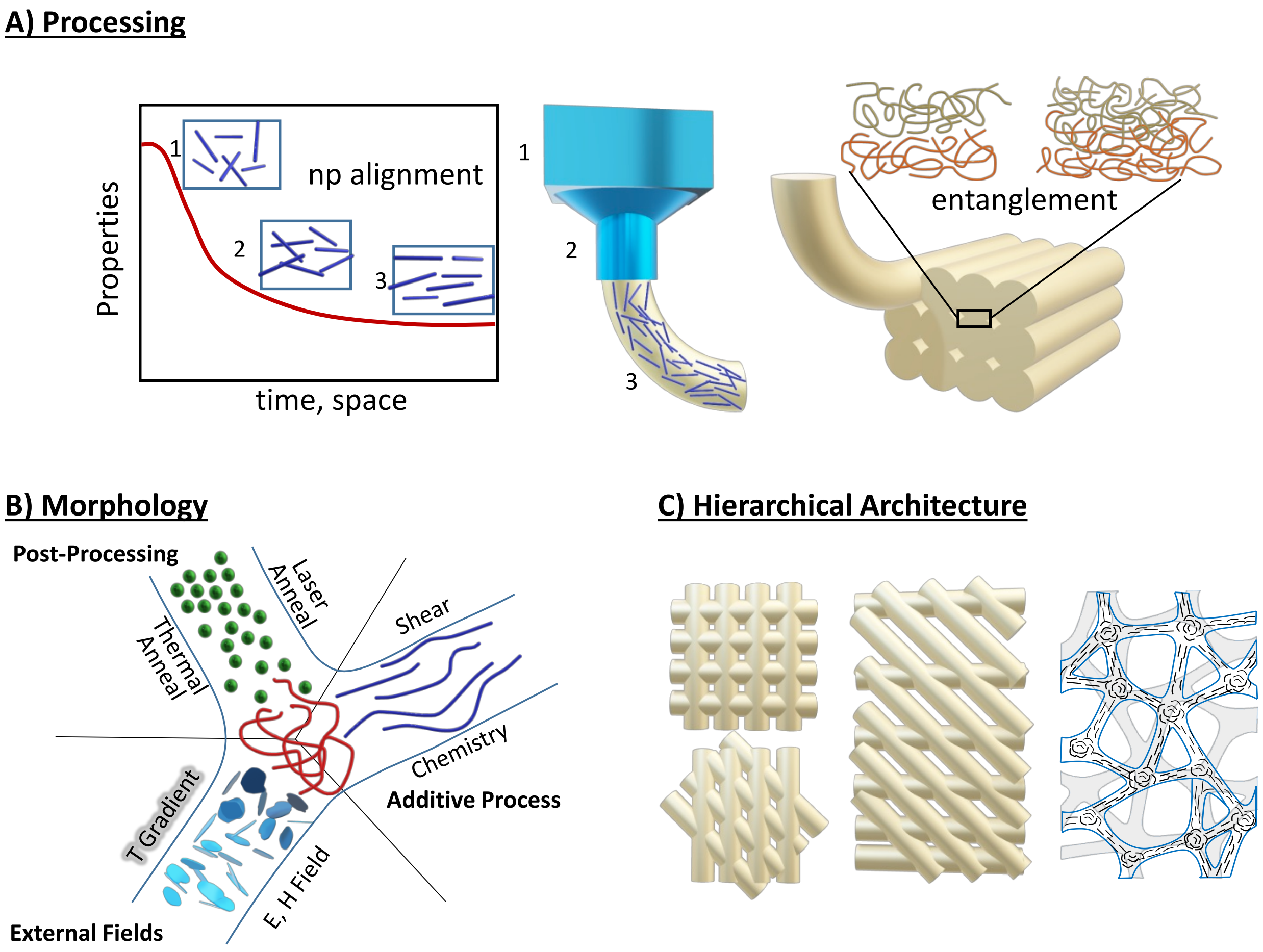

**Figure 2**: Overall schematic of areas within additive manufacturing in which nanoparticles can play a role as enabler for AM or be used to enhance functionality (mechanical, electrical, thermal etc.). A) Processing, represented by the fused deposition modeling process which includes extrusion, melt flow, road-to-road bonding (welding, coalescence, and entanglement), shape evolution. Spatial and temporal evolution of nanoparticle alignment and assembly dictate morphology and performance of printed parts. Road-to-road adhesion is improved by entanglement or crosslinking. Properties that evolve over time are thermal, electrical conductivity, thermal coefficient of expansion, strength, modulus, density, heat capacity, crystallinity and other. B) Morphology is governed by the additive manufacturing process with external aids such as magnetic (H), electric fields (E), temperature gradients or additives that influence the chemistry and kinetics during printing. C) Architecture is dictated by hierarchy from tool path topology of the design and alignment of sub-elements within the roads and layers of the additive manufacturing process, such as polymer chains, nanoparticles and larger constituents, such as reinforcement fillers.

While it is more obvious to imagine how nanoparticles could aid in the additive manufacturing process, it may also be useful to look at how additive manufacturing offers opportunities to advance nanocomposites. The advancement of nanocomposites requires a synergistic effect, which may include voxelated morphologies enabled by nano-elements that allow discrete, localized property manipulation to address certain functions of the printed part (fold, reinforcement, RF response, to name a few).[37] In general, the ability to build complex hierarchical structures and assemblies with nano-elements that are otherwise not obtainable through conventional processing leads to true advances in nanocomposites.

We tried to capture all areas that will positively affect processing, morphology, and architecture and lead to overall improvements in performance of printed parts beyond what is possible without the combination of AM and nano. In evaluating and selecting research efforts in this area, the selection criteria was based on the relationship between AM and the nanofiller used, as opposed to organizing by certain classes of materials. We did not include new materials development for either additive manufacturing or nanocomposites or a combination of the two. These are important areas that still need to be addressed.[40] Off-the-shelf, commodity materials are good for the most rudimentary additive manufacturing processes, which include complex shaped tools or prototyping for fit, form and function. However, true advantages will only be enabled when new matrix materials are developed that adapt to the processing conditions, such as FDM or SLS. Several groups use nanoparticles that both improve or enable processing, while also contributing functionality in the printed part. Others have utilized nanofillers to aid in tailoring the morphology of the nanocomposites and architecture of manufactured parts.

A combination of both nanocomposite enabled additive manufacturing and additive manufacturing enabled nanocomposites is achievable with strategic choice of nanoelements. Individually, the matrix and nanoparticle reinforcements can be selected for an application based on their properties, while selecting an AM process that enhances the end products that are created, which would not be possible with traditional manufacturing methods. Full utilization of the relationship between these two technologies has the potential to unlock opportunities for advancement in multiple facets, such as biomedical, vehicular transportation, and satellites in space. The potential for the synergy harnessed by the following research is still largely unexplored and holds much promise for the future. The following chapters will detail the innovations in each area-processing, morphology, and architecture-highlighting the notable advances in technology that show much promise for the future relationship of AM and Nano.

## 2 AM enabling nanocomposite morphology control | Nanocomposites enabling AM

### 2.1 Process Control

Critical to optimal printability of low viscosity, liquid nanocomposite dispersions into stable solid structures as the material exits the nozzle is the comprehensive understanding of the rheological behavior during the entire process. This includes the consideration of geometrical flow conditions, melt pool features and characterization of viscosity at each step to capture shear thinning and gelation characteristics of each material as a function of composition and printing parameters (temperature, printing speed, flow rate, pressure etc.). In addition, the effect of the AM process enabling filler on the matrix solidifying process, such as photopolymerization, crystallization, and vitrification, have to be studied to control emerging properties of the printed parts (local stresses, interfaces/phases, CTE gradients), similar to what has been accomplished in e.g. injection molded nanocomposites.[41] The materials discussed throughout this chapter are selected based on their ability to interact in a synergistic manner with the respective AM process for a given application. The variety of materials is vast to encompass the potential for growth in this and adjacent areas using AM and Nano.

### 2.1.1 Self-support and shear thinning

One of the first challenges for polymer-based additive manufacturing is the processability of polymers and polymer nanocomposites and the requirement that deposited material changes from a flowable state into a solid, self-supporting state and ensure proper welding of materials to existing layers. Self-support needs to furthermore extend to deposition of material of many more layers. Overhanging features and thin-walled structures still require support material to address gravitational sagging. Self-support can be accomplished via crosslinking a liquid resin material locally or by dropping temperatures from above the melting temperature or glass transition to a crystalline or glassy state. Highly filled pastes can be printed without material flowing, however, postprocessing will lead to a collapse of the printed features.[42, 43] This is a typical process in binder type additive manufacturing processes in which the binder material is removed with substantial shrinkage of the remaining typically ceramic materials.[18]

For materials that remain liquid at room temperature and cannot be UV-cured during extrusion based direct write processes, rheology modifiers are added to obtain gel formation or thixotropic material properties that allow processing and self-support after the material is deposited.

Printability of materials is governed by properties associated with material flow, primarily the shear thinning behavior of Bingham plastics, which after cessation of flow return to their higher viscosity in the quiescent state with a minimum yield strength that allows the material to behave solid like and not collapse under its own weight or the weight of successive layers.

This methodology can be extended to materials with viscosities that are too high for processing at or above room temperature. In this case, low volume fractions of a solvent can be added to the formulation with rheology modifier to reduce viscosity or other passive additives (active during processing with no additional functionality in the final printed part) such as nano-sized layered silicates and fumed silica can be added to control the rheology (**Figure 3**).[44-46]

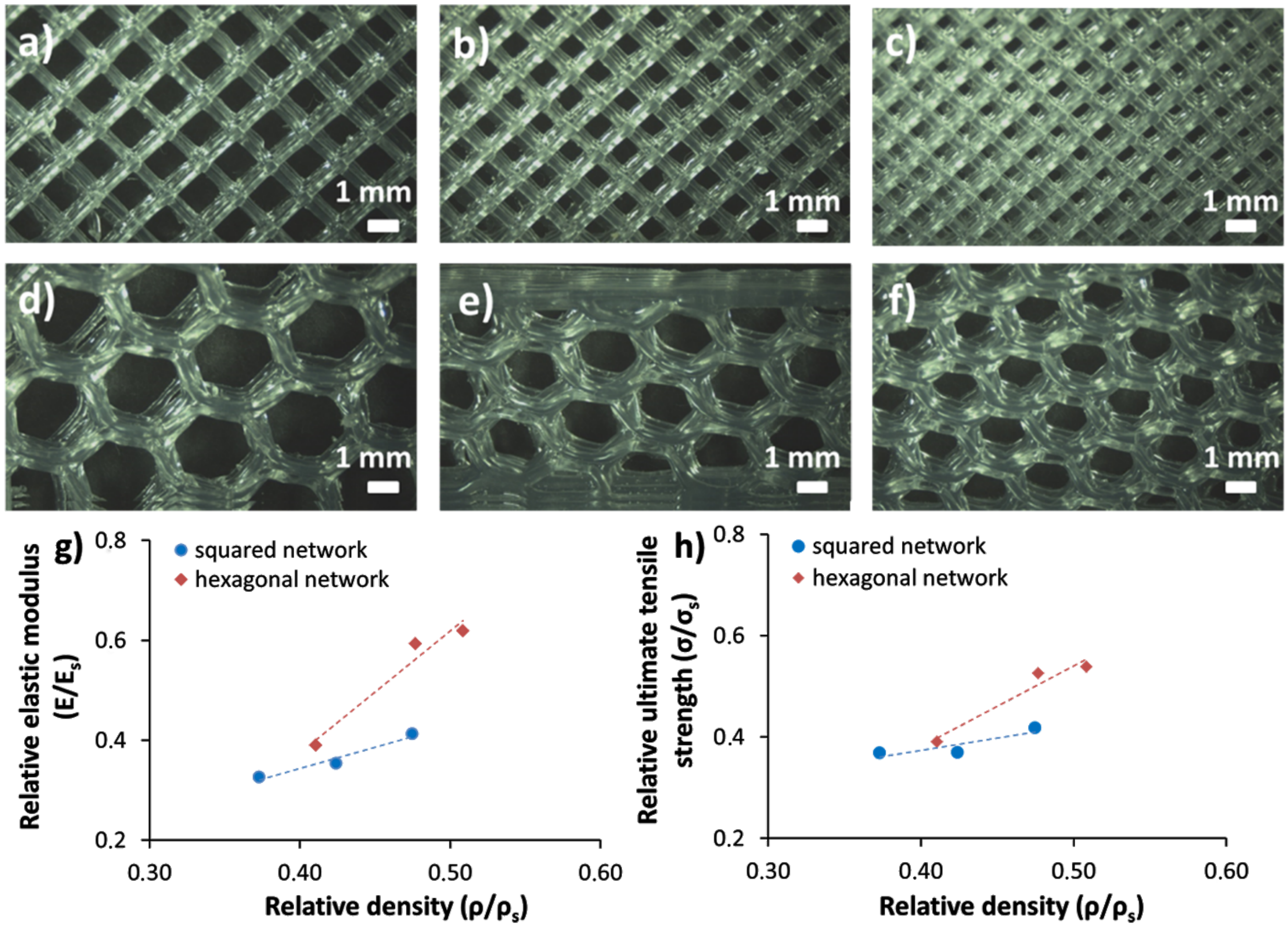

**Figure 3**: Optical stereoscopic images of the UV-3D printed specimens with (a–c) squared unit cell and (d–f) hexagonal unit cell at increasing infill density. Plots of (g) the relative elastic modulus and (h) relative ultimate tensile strength as a function of the relative density of the specimens (dotted lines are a guide for the eye).[45]

Compton et al developed an epoxy based ink with layered silicate nanofiller at moderately high concentrations (10-15 vol%) to induce gelation of the normally liquid resin at room temperature, while also allowing for shear thinning during processing.[47] Addition of the nano-filler allowed tailoring of the composition and rheology with additional high aspect ratio micron-sized reinforcement fillers to additively manufacture hierarchical structures such as cellular structures. One disadvantage with organically modified layered silicates as a rheology modifier is the addition of 25-30% low molecular weight surfactant molecules (typically long alkyl chain ammonium salts) that will negatively affect thermal and mechanical performance of resulting structures. A similar approach is used to control the viscosity of biocompatible hydrogels in poly (ethylene) glycol-alginate. A biocompatible nanoclay (Laponite XLG) was added to the formulation to obtain shear thinning behavior and setting of the hydrogels once exiting the print nozzle. Small cellular structures were printed using a syringe pump system (Fab@Home) that were found to be cell-friendly and stretchable.[48] A few other articles discuss process modeling efforts to understand the additive manufacturing process with nanoparticles. Gu et. al describe a multiscale computational numerical modeling framework that may become an important tool in predicting thermodynamic and kinetic mechanisms during SLM AM/3DP processes with nanoparticle contributions.[49]

Additionally, in the bioelectronics sector, hexagonal boron nitride (hBN) polymer nanocomposites offer lightweight weight, electrical insulating, viscosity modifying, and thermally conductive attributes which make them an attractive material for printing tailored architectures without sagging or deformation. Poly(lactic-co-glycolic acid) (PLGA) and hBN particles are dispersed in a trisolvent system that consists of dibutyl phthalate (DBP), ethylene glycol monobutyl ether (EGBE) and dicholoromethane (DCM) to create an ink for printing of complex geometries.[50] The trisolvent system was customized to include properties of a plasticizer, a surfactant for dispersion of hBN particles, a high volatility solvent to dissolve the PLGA, and a mechanism for solidification during the printing process. This direct write manufacturing method with custom PLGA-hBN ink allows for freedom to control various properties such as porosity, stiffness, tensile strength, and thermal conductivity by altering the volume fraction of hBN used in the nanocomposite (**Figure 4**).

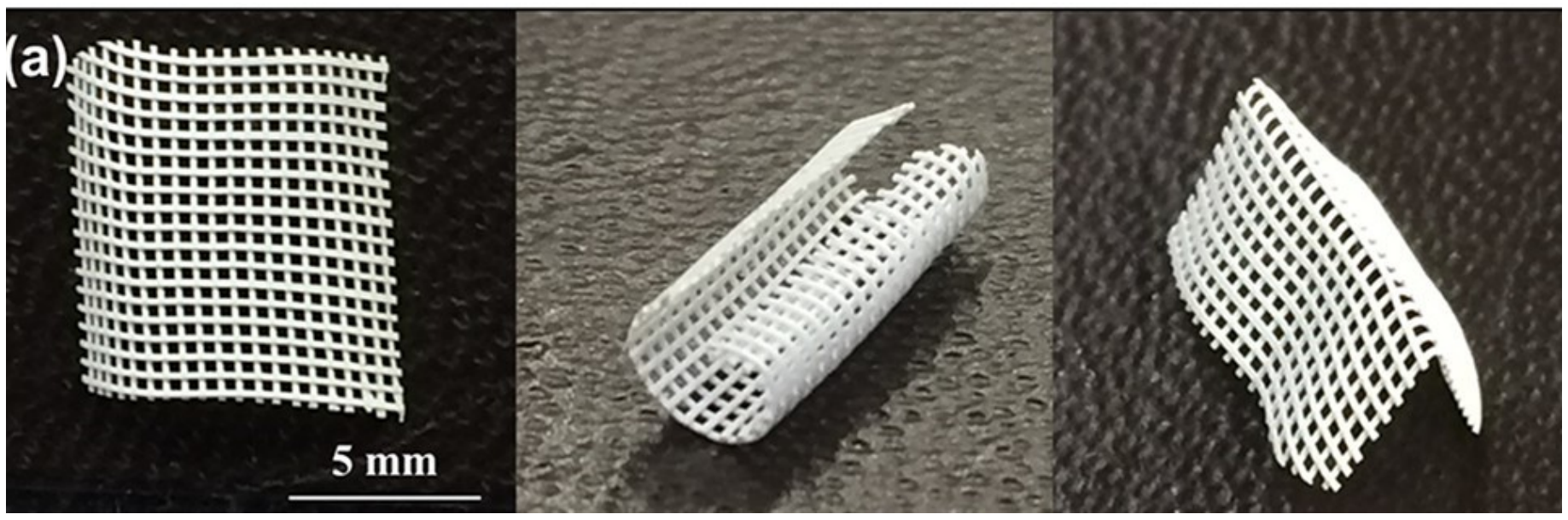

**Figure 4**. Mechanical properties of 3D printed hBN; photographs of a 3D printed 40% vol hBN gridded sheet that has been cut, rolled, and folded but still maintains its mechanical integrity[50]

In a unique way, Huang et. al demonstrated printing of liquid materials via their 'printing-then-solidification' process. The fundamental process is similar to printing with clay as a rheology modifier, however, in this case the clay is not added to the ink but the medium the printhead prints into. A solution of synthetic clay (Laponite) is used as a bath. The printhead deposits liquid materials to the bottom surface of the bath and the thixotropic behavior of the bath solution allows printing when the head moves, while the gelation of the solution in the quiescent state will arrest the liquid in its printed state immediately after the head moves on (**Figure 5**). The authors also demonstrate that this enables printing of otherwise unprintable resins to be printed in the z-direction.[51]

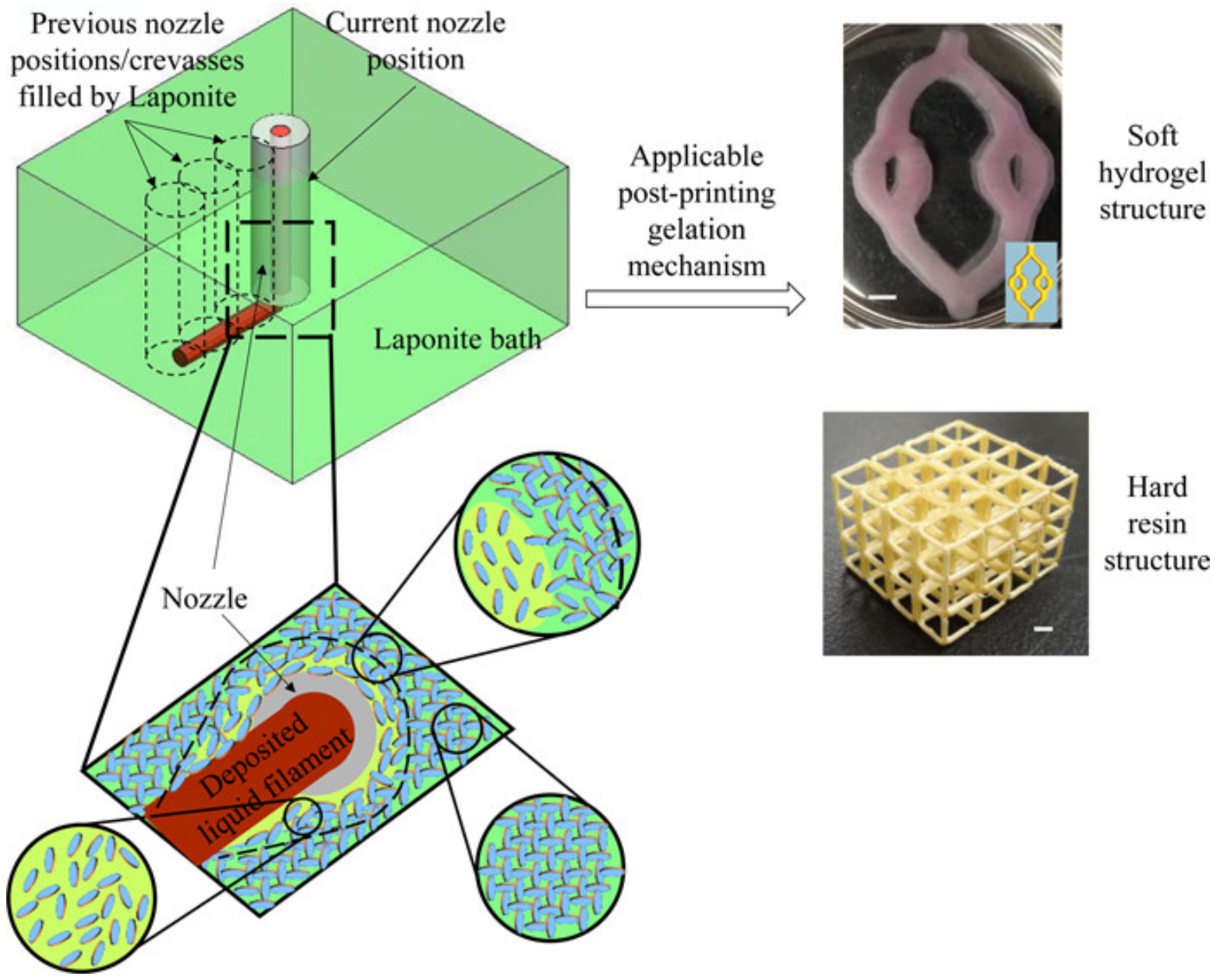

**Figure 5**: Using the printing-then-solidification approach, the printed structure remains liquid and retains its shape with the help of the Laponite support bath. Then the completed liquid structures are solidified in situ by applying suitable cross-linking mechanisms. Finally, the solidified structures are harvested from the Laponite nanoclay support bath for any further processing as needed.[51]

The utilization of nanoparticles to improve the useability of hydrogels in an AM setting has been observed through various research efforts. Thermo-responsive poly (urethane) (PU) nanoparticles based on mixed oligodiols have been used to form gels. These systems gel at temperatures below 37$^{o}$C without a crosslinker, and stiffness can be matched to neural tissue by changing the ratio of the two PU nanoparticle concentrations. Additive manufacturing of the gels was accomplished using fused deposition modeling type printing.[52] Unterman et. al used nanosized platelet particles with distinctly different aspect ratios which allowed them to independently control rheology (lower aspect ratio Laponite) and mechanical properties (larger aspect ratio Montmorillonite).[53] By tuning the nanoplatelet filler content, the exfoliation of the nanoplatelets enables users to control the solution flow, which aids in facilitating injection and manipulation into a DIW style AM process. This work is ideal for the advancement of drug delivery systems that must be precisely tailored.

Shear thinning has also been explored using cellulose nanofibrils (CNFs) that form hydrogels in aqueous solutions.[54] Wang et al. employ CNF/polymer filament writing at different composition to create gradient films. Once the filament is extruded to the build plate, a self-healing process of the roads leads to coherent films with gradient patterns of choice. In other work, direct write of aqueous hydrogels of CNFs is accomplished through the shear thinning and recovery properties of the nano-enabled gels. These techniques are mostly used for printing biomaterials and recyclable and renewable materials or carry other functional nanofillers.[55-60] CNFs are typically used in combination with the fast cross-linking ability of alginate to obtain noncytotoxic formulations that can be printed into living soft tissue with cells such as anatomically shaped cartilage structures (ear, meniscus).[60] Wessling et. al have demonstrated a laser-less additive manufacturing route towards membrane electrode assemblies using a blend of CNCs as a binder and rheology modifier in high solid content metal/ceramic (titanium, silicon dioxide and aluminum oxide) pastes.[61]

To control a wide spectrum of flow responses from Newtonian to viscoelastic percolating network forming, fumed silica has also been utilized as a rheology modifier with primary particle sizes of 5-30nm in photocurable poly(urethane) acrylates that contain alumina platelets for reinforcement and hierarchical morphology with magnetically assisted processing.[62] The hierarchical morphology of locally aligned platelets enables to print overhanging features during direct write processes of composite materials without using support material.

Aside from layered silicates, attapulgite nanoparticles were successfully used to control the rheology of the stereolithography resins[63]. Peptides that self-assemble into ribbons, sheets and fibrils have also been explored for direct write processing bio-compatible scaffolds (**Figure 6**).[64]] The bottom-up approach towards 3D scaffolds takes advantage of their biomimicry, stimuli-responsive properties, biocompatibility/degradability, and functionalization.

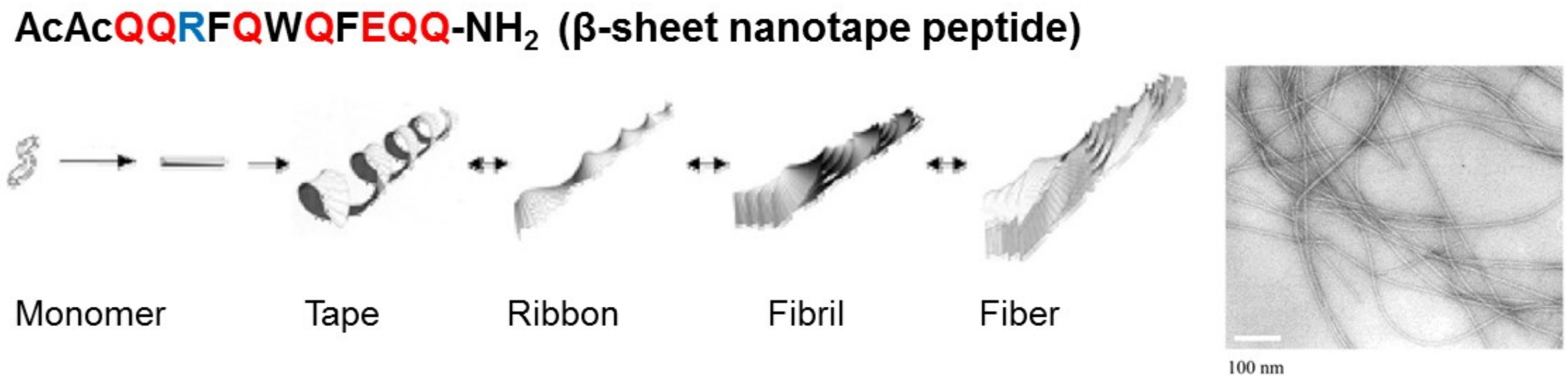

**Figure 6**: Self-assembly of short peptides yields nano-fibrous hydrogel scaffolds. The macromolecular assemblies are stabilized by overlapping hydrophobic interactions.[64]

A narrow window in the phase diagram of an alumina nanosol showed the existence of well-defined regions for formation of stable sol, reversible gel, and irreversible gels. The reversible gelling behavior showed yield stress depending on interactions and exhibited a drop-in viscosity by 6 orders of magnitude due to the formation and breaking of hydrogen bonds between octahedral alumina moieties. Multilayer structures using direct write and ink-jet techniques were demonstrated.[65]

Selective laser sintering (SLS) requires controlled size distribution of powder particles to enable optimal flowability during the melt process. To improve flowability, nanoparticles (fumed silica) were dry coated onto micron-sized (few to 10s of microns) polymer particles (polystyrene, polybutylene terephthalate) by wet-grinding at low temperatures.[66] To address the remaining interstitial spaces between micron and sub-micron sized powder particulates, which lead to substantial part shrinkage and porosity during consolidation, nanoparticle additives are suggested for improving the sintering process. If these nanoparticles are made from metals, further enhancement in the laser sintering process can be accomplished due to the improved energy transfer during laser processing between the different types of particles.[67]

The rheological modifiers used in the studies discussed exhibit an ability to overcome some obstacles that are typically seen in various AM applications, especially when additional support is needed. Though the increased viscosity in the nanocomposite during deposition administers the structure needed for single roads and layers with enough stiffness to maintain their structural integrity, it does present another issue. The ability for the single roads or adjacent layers to diffuse together is minimized, which could be undesirable for certain applications that require high strength in the out of plane direction. Of course, a post processing method such as annealing could assist in addressing these concerns.

### 2.1.2 Energy transfer, welding

While anisotropy both on the road level (alignment of constituents) and on the part level (alignment of roads) is an opportunity in AM, a typical challenge in the layer-by-layer process during most additive manufacturing is that the mechanical properties in the normal direction to the layers in this 2.5D process is significantly reduced (<75%) compared to the in-plane properties. This is because adjacent road-to-road deposits are still within a temperature difference that allows for entanglement and diffusion of polymer chains for good road adhesion. Once the process starts depositing the next layer, the bonding quality is reduced and flaws in the z-direction accumulate to produce anisotropic mechanical properties of the final part. Because nanoparticles are used as processing aids to overcome some of these challenges, energy transfer and welding of layers and roads is covered under the Process Control section.

Sweeney et al. applied locally Induced Radio Frequency (LIRF) welding that uses local heating properties of nano-elements at the surfaces of printed filaments to enable additive manufacturing of high strength materials.[68] Conventional thermoplastic filaments are coated with multi-wall carbon nanotubes in a bath batch process. Fused deposition modeling of these filaments led to a macroscopic structure with MWCNTs at the interfaces between printed roads that can be selectively heated by exposure to microwave irradiation (**Figure 7**). This in turn leads to enhanced mobility and entanglement at the interfaces and overcomes typical z-directional losses in mechanical properties. The authors demonstrate overall improvements in strength of the final 3D printed part with isotropic mechanical properties of fused filament part that exceed conventionally manufactured counterparts.[69]

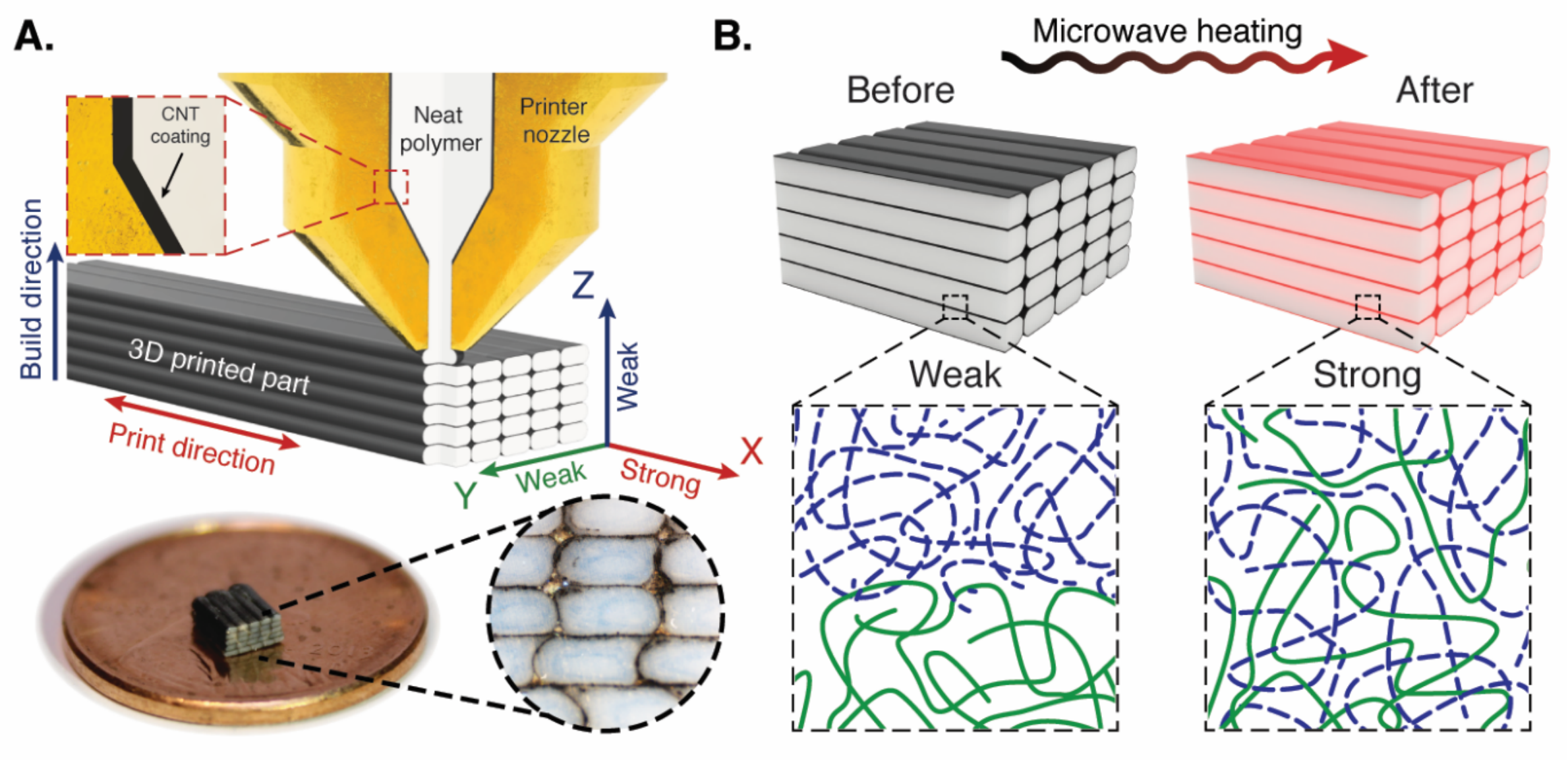

**Figure 7**: (**A**) 3D-printed parts tend to display weak tensile properties in the *y* and *z* directions due to poor interlayer welding. To address this, we coated thermoplastic filament with a CNT-rich layer; the resulting 3D-printed part

contains RF-sensitive nanofillers localized at the interface. (**B**) When a microwave field is applied, the interface is locally heated to allow for polymer diffusion and increased fracture strength.[69]

In other inductive heating techniques, such as the patent by Champion and Abbott, a combination of nanoparticle susceptors are added to enable heating of feedstock material at specific stages of the AM process.[70] Susceptors can be designed to have specific absorbance to microwave radiation that allows directed and selective energy absorption of only one nanoparticle while additional nanoparticles can be triggered at other frequencies. In addition, at least one of the susceptor particles may decompose at higher temperatures that are induced by the second stage susceptor. Kim et. al study the thermodynamic behavior of powders during laser heating and apply molecular dynamics simulations to study Ag nanoclusters (2-7nm) during heating and cooling process and find a mesoscale regime where properties of nanoclusters can be described with macroscopic concepts.[71] In another patent,[72] magnetic nanoparticles are added to the feedstock to accomplish heating via high flux, frequency sweeping, and alternating magnetic field in the nozzle assembly.[72] While Erb et al. use magnetic fields to obtain improved morphology and mechanical properties of resulting parts, the decoration of larger size anisotropic filler particles with magnetic nanoparticles leads to alignment and therefore shear-thinning of formulations with higher filler loadings.[73]

Improvements of energy transfer have also been accomplished in additive sintering techniques. Gold nanoparticles are used to increase in coupling of laser energy and the material to be sintered (e.g., semiconductors such as ZnO). In one example, Au nanoparticles are absorbed to crystalline ZnO and are exposed to laser radiation thereby leading to sintered metal-semiconductor hybrid materials with potential in light harvesting applications.[74] The sintering process was improved using silver nanoparticles in the light sintering of aerosol-jet printed silver nanoparticle inks on polymer substrates and led to much greater performance in electrical properties over thermal sintering techniques. In this case, a Xenon light source was used to induce inductive heating via the presence of the nanoparticles.[75]

Magdassi et al. use semiconductor metal hybrid metal nanoparticles (37x4nm CdS with a 1.5nm Au tip) which form radicals through a photocatalytic process that lead to charge separation after light absorption and electron transfer to the metal tip. This enables redox reactions to form radicals under aerobic conditions. The authors demonstrate 3D printing under water with substantially faster and more complete photopolymerization (**Figure 8**).[76]

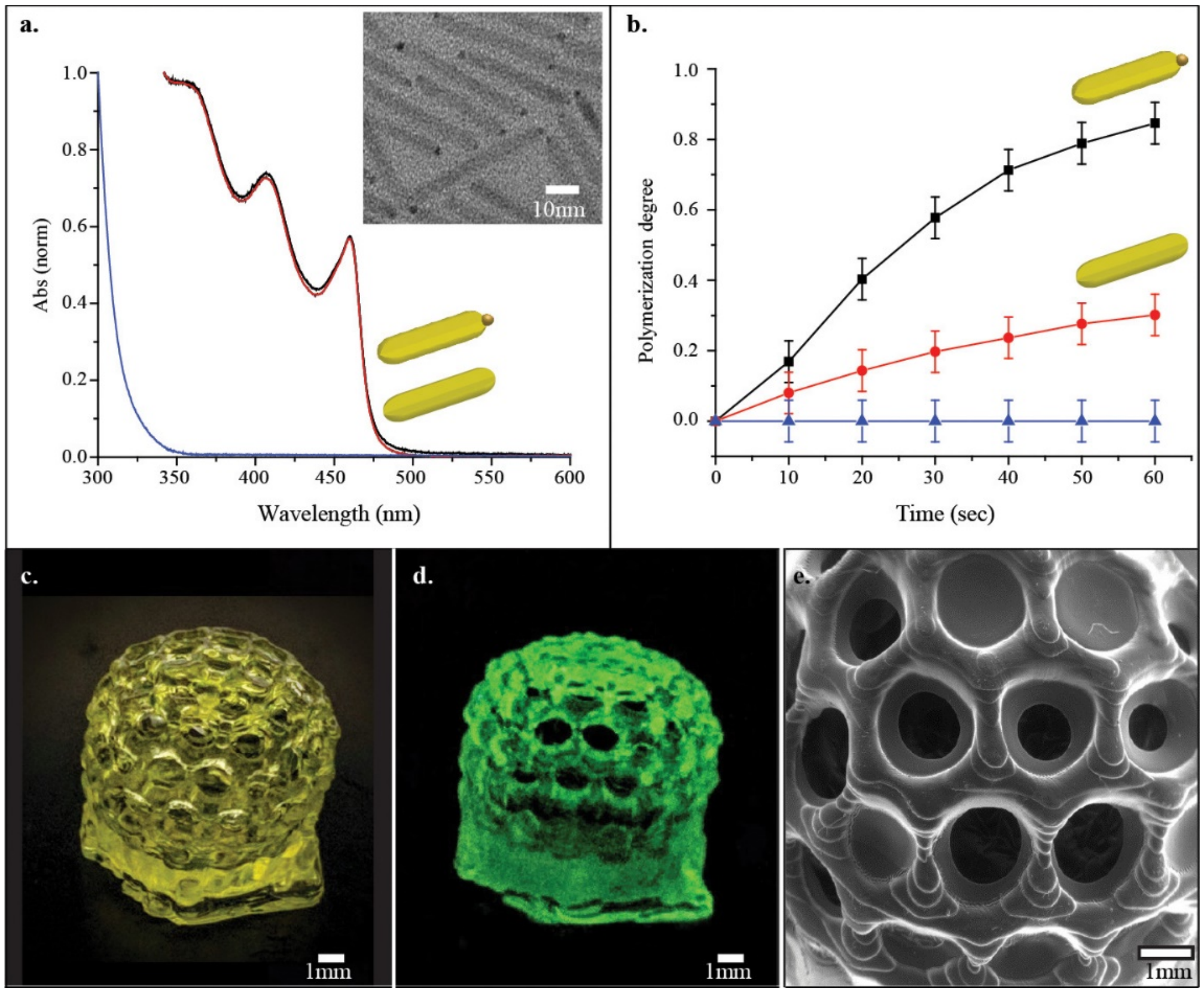

**Figure 8**. The 3D printing using HNPs as photoinitiators. (a) Absorption spectra of CdS–Au HNPs (black), bare CdS rods (red) and I2959 (blue). Inset: TEM image of CdS–Au having rod dimensions of 37 × 4 nm (length × diameter) with 1.5 nm diameter gold tip. (b) Polymerization degree under UV light at 385 nm with intensity of 20 mW/cm$^2$ using CdS–Au (black), CdS (red), and I2959 (blue) as photoinitiators. The CdS–Au and CdS are in the same concentration (0.4 μM), whereas the I2959 photoinitiator is in its optimized concentration for polymerization (0.02 M). The polymerization degree of acrylamide was calculated from the decrease of the Fourier transform infrared (FTIR) absorption peaks of methylene group vibrations at 988 $cm^{-1}$ (assigned to the out-of-plane bending mode of the ═C–H unit) normalized to the C═O stretch peak at 1654 $cm^{-1}$ as an internal standard. The error bars are assigned to be the maximum error as extracted from triplicate measurements. (c–e) Images of a 3D printed Buckyball using CdS–Au HNPs as the photoinitiators and CdSe/CdS seeded nanorods as fluorescent markers. (c) Regular light photo. (d) Fluorescence image under 365 nm excitation. (e) Scanning electron microscopy image of the dried structure.[76]

Similarly, a photon up-conversion in rare-earth-doped luminescent materials enabled the use of cost-effective and commercial low-power continuous wave near-infrared laser diodes in additive manufacturing of light sensitive resins that contain photo-initiators for radical polymerization. Outstanding, high-intensity near IR to UV blue-up-conversion emissions from $Tm^{3+}$ doped $K_2YbF_5$ micro- and nano-crystals was accomplished.[77]

In another study that harnesses the power of lasers, Lin et al. utilizes a hybrid additive manufacturing process that combines laser deposition of graphene-metal nanocomposites and laser shock peening (LSP) on nearly every printed layer.[78] The printed parts exhibit improved fatigue performance as a result. Each layer of the graphene-iron nanocomposite is laser sintered before LSP is introduced for improved mechanical properties. The introduced shock loading creates nanotwins-a sort of linear boundary in the metal’s microstructure that are formed from the large stress around the graphene-iron interface. During the LSP process, the graphene nanofillers allow the wave from the shock peening to bounce back and forth between the nanotwins and grain boundaries in the nanocomposite instead of constraining the motion and increasing residual stress. The waves incurred in this process subsequently create nanowrinkles in the graphene, which were shown to reduce the crack propagation rate in the material. This energy transfer process improves the SLS produced graphene-iron nanocomposite with mechanical properties that would not have been achieved otherwise.

### 2.1.3 Improving reaction kinetics

It is well established that the addition of nanoparticles to reactions such as the crosslinking chemistries of thermosets can alter the kinetics of the reaction or the crystallization of thermoplastic materials.[79-81] In addition, the presence of anisotropic (layered silicates) or chaining (carbon black) nanoparticles can induce preferential crystallite orientation with respect to flow direction. The same principles underlying conventional extrusion and injection molding processes apply to the additive manufacturing process.

While neat nanoparticles can, in some instances, induce changes in the kinetics via catalytic effects, it is common to functionalize the surface of nanoparticles to make them compatible with the matrix polymer.[79, 82] For example, compatibilizing surface ligands such as silanes on ZnO nanoparticles is thought to considerably increase conversion in the crosslink reaction in the UV photopolymerization process during inkjet printing, thereby leading to additively manufactured nanocomposites with improved thermal and mechanical properties.[83]

Photopolymerization kinetics were also improved significantly by using water-dispersible photoinitiator nanoparticles based on 2,4,6-trimethylbenzoyl-diphenylphosphine oxide (TPO). The TPO nanoparticles enhance the reaction kinetics in the range from 385 to 420 nm. This allows printing that is otherwise impossible without adding solvents in commercially available, low-cost, light-emitting diode–based 3D printers using digital light processing.[84]

The inclusion of neat oxide nanoparticles without modified surface into preceramic polymers for the formation of oxygen containing advanced ceramics has shown favorable reaction kinetics and enables the production of phase pure ceramics at low temperatures. In this case, the nanofiller reacts with decomposition gases during pyrolysis or ceramic residues from preceramic polymers that are printed using paste extrusion to give single phase ceramics with the potential of creating crack-free bulk components.[17]

While reaction kinetics are improved in these examples, the presence of nanoparticles can also affect the reaction negatively as demonstrated with organically modified montmorillonite and attapulgite.[63] Though the addition of nano $SiO_2$ increased the curing speed of the SLR matrix, the use of OMMT and ATP decreased the curing speed and did not improve the mechanical properties or dimensional accuracy of the finished parts from that of current commodity stock materials.

Wang et. al reported on hybrid nanosized photoinitiators with low cytotoxicity and migration by coupling of polyhedral oligomeric silsesquioxanes (POSS) to benzophenone derivatives with uniform sizes, low tendency to migrate, robust resin viscosity, enhanced thermal stability and mechanical strength, increased photoactivity, and lower cell toxicity compared to their corresponding benzophenone molecules.[85]

When compared to polymer materials, only a limited number of metals and metal alloys can be printed today. Martin et. al demonstrated that these issues can be resolved by introducing nanoparticles of nucleants that control solidification during additive manufacturing. The authors chose nucleants that are compatible with 7075 and 6061 series aluminum alloy powders and found that these previously unprintable high-strength aluminum alloys could now be processed successfully using selective laser melting (**Figure 9**).[86]

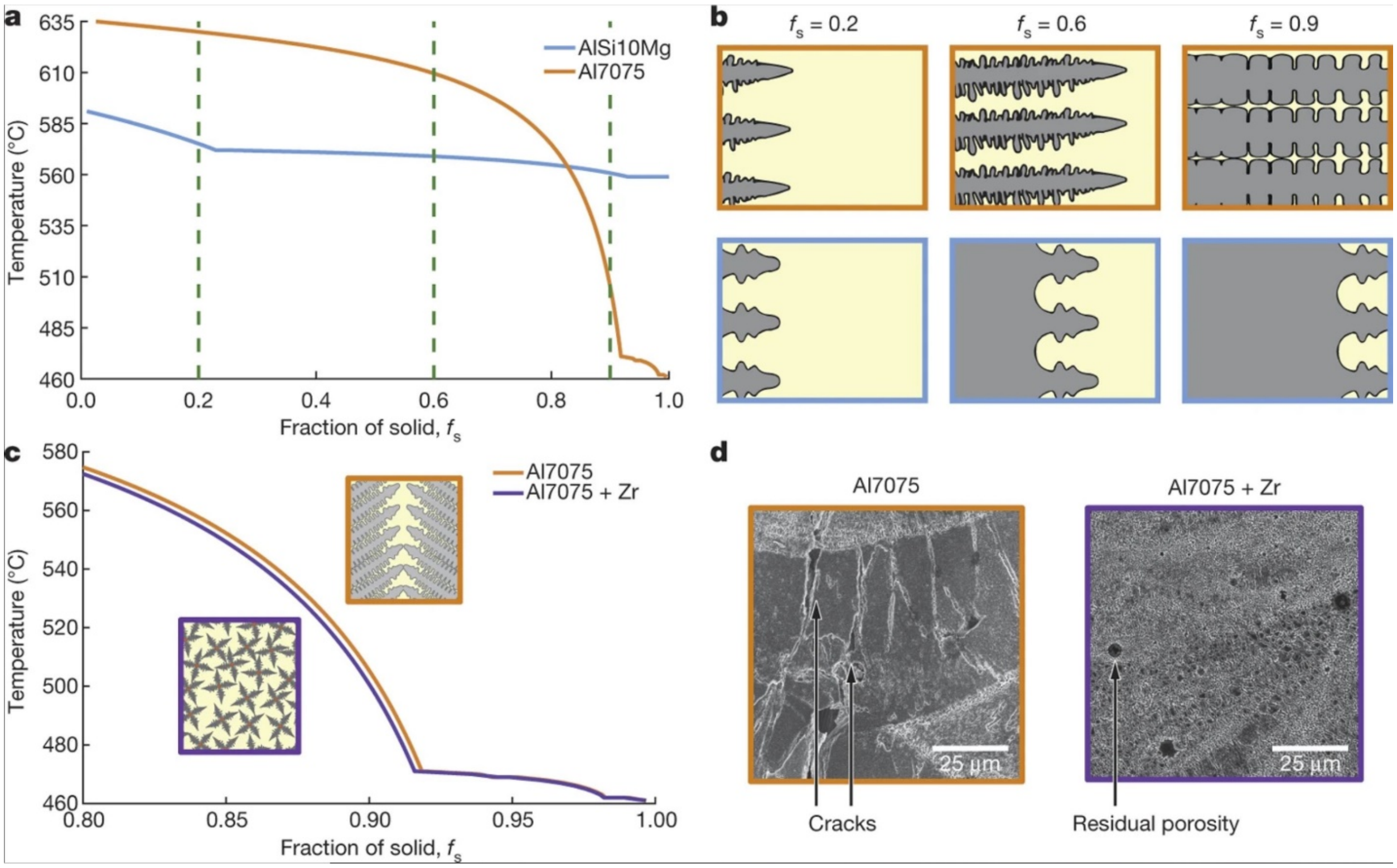

**Figure 9**: a, Solidification curves for Al7075 (orange) and conventional 3D-printed aluminum (AlSi10Mg; blue). b, Schematic representation of solidification, indicating how solidification over a large temperature range leads to long channels of interdendritic fluid that result in cracking (top; such as in Al7075), whereas a small solidification range leads to short interdendritic regions that can easily be backfilled (bottom; such as in AlSi10Mg). The three panels in each row correspond to the solid fractions indicated by the vertical green dashed lined in a. c, Adding zirconium to Al7075 (purple) has little effect on the solidification behavior at high solid fractions, at which alloys are the most tear- and crack-susceptible. d, Polished and etched scanning electron microscopy (SEM) images depicting the resulting microstructures with (right) and without (left) the addition of zirconium.[86]

### 2.1.4 In situ formation of nanoparticles during printing

While adding nanoparticles to the feedstock resin is following known recipes from conventional nanocomposite processing routes, some research groups are also exploring the formation of nanoparticles during the AM process. This can eliminate the problem of dispersing nanoparticles in the feedstock material and may improve interfacial bonding between matrix and particles in the product. In a dedicated sol-gel treatment step following a digital light process (DLP) with photo-curable inks, nanoparticles were formed in situ, thereby eliminating the often-difficult incorporation of inorganic powders into ink formulations. The formed nanoparticle phase is covalently bonded to the organic matrix by using methacrylate-based silanes as organic-inorganic bridging moieties and tetraethoxysilane as inorganic precursor forming silica nanoparticles. The resulting materials exhibit an increase in the breadth of loss tangent, an increase in the $T_g$ by up to 15$^o$C, as well as a doubling of the storage modulus of printed coupons.[87] The authors speculate that covalently bonded nanoparticles form a separate network to enhance overall properties.

In a related process, nano-sized glass powder for sintering applications was produced via sol-gel reaction of a precursor solution and heat-treatment.[88] Such processes present useful ways of producing controlled morphology on the fly during the printing process in both direct write processes and sintering applications. Borrowing knowledge from the conventional photography development process, silver nanoparticles were produced in situ during photo-crosslinking (DLP) of silver salts ($AgNO_3$) containing formulations for the additive manufacturing of micrometer scale complex electromechanical parts. In situ formation of nanoparticles enables good dispersion with some control of crosslinking the matrix simultaneously.[89]

While the primary scaffolds were printed via extrusion based processes, the scaffolds were coated with Ca-P nanoparticles in a separate post process sintering step which resulted in the formation of uniform self-assembled Ca-P/polydopamine nanolayers on the strut surface of scaffolds.[90] The photothermal effect of the scaffolds in this study proved to significantly decrease the cell viability of various cancer cells and effectively inhibited tumor growth in test subjects as well.

Simultaneous two-photon polymerization using composites containing Au nanoparticle lead to highly complex structures on the micron-scale.[91] A near-infrared (780 nm) femtosecond laser beam is used to initiate the photoinitiator within the resin by the simultaneous absorption of two photons that subsequently triggers local chemical reactions, including monomer polymerization, cross linking and Au metal salt precursor reduction. This leads to simultaneous formation of both the polymer matrix and metal nanoparticles, while the in situ generated nanoparticles were instantaneously embedded in the polymer matrix which prevented further growth.

Photo-curable liquid formulations containing different amounts and types of silver salts as precursors to silver nanoparticles (AgNP) were used in stereolithography (SLA) to create a homogenous nanocomposite with antibacterial properties.[92] The laser light simultaneously activated the cross-linking of the acrylic resin and reduced the silver cations into metallic AgNPs. In this way, the silver acrylate and silver methacrylate salts become an integral element of the cross-linked, homogenous structure and increases the effectiveness of the interface interactions between the polymer matrix and nanofillers. The SLA process coupled with the presence of the in situ generated AgNPs allow for higher precision and custom geometries to be created.

Similarly, in SLM printing of aluminum matrix composites, CNT is produced in situ within aluminum powder in a study by Geng et al.[93] To perform the in situ impregnation of CNT on the aluminum powder, nickel catalysts were uniformly electroplated on the powder prior to the SLM print. Fluidized bed chemical vapor deposition (FBCVD) is used to expose each particle to a $C_2H_4$-$H_2$-Ar gas mixture for synthesis of CNT without changing the original morphology of the Al powder. Following the preparation of the CNT within the aluminum powder, the improved raw material is then printed using SLM. The presence of the in situ synthesized CNTs reduced the reflectivity of the laser by 11%, making geometrical control of the parts more facile. Additionally, the lower reflectivity manufacturing produced a smooth surface finish with little to no defects in the printed part.

Ceramic matrix composites harness the ability to create in situ nanoscaled architectures through the selective laser melting process of microsized Ti-6Al-4V and $TiB_2$ powders.[94] During the AM process, the $TiB_2$ particles are transformed into a needle-like, nanoscaled TiB that are rich in some regions and less populous in others. With larger concentrations of $TiB_2$, aggregation begins to occur, and wear loss increases. Despite the absence of cracking, the micrographs of the finished nanocomposites exhibited randomly distributed microspores around the area that surrounded the molten pool (**Figure 10**). Although these defects are caused by insufficient time for the molten liquid pool to merge with the previously melted layer, the $TiB_2$ particles cause the melt pool to have larger width and depth. The larger dimensions of the melt pool help to reduce defects from this portion of the AM process leading to full density.

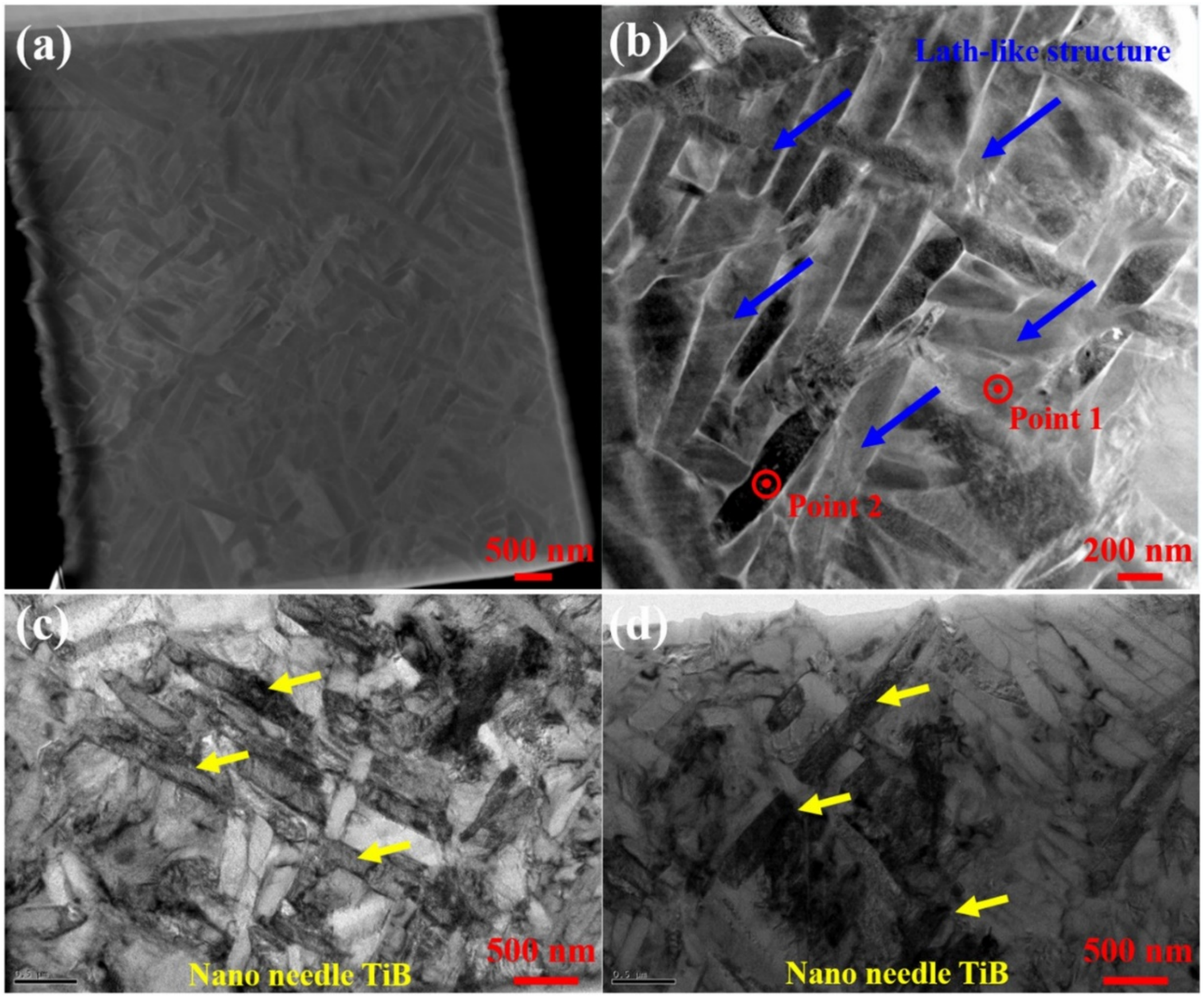

**Figure 10** (a)The integrality dark-field morphology of the TEM foil extracted via FIB, (b)high magnification dark-field micrographs, (c)and (d)high magnification bright-field images.[94]

### 2.1.5 Surface finish, dimensional tolerance

The layer-by-layer methodology in many of the additive manufacturing processes leads to build up of thermal gradients within the printed part due to thermal fluctuations within the build chamber or the build platform. In addition, temperature fluctuations at the print head may lead to non-uniform material deposition including variations in road diameter during FDM or layer thickness during SLS. These effects lead to non-uniform thicknesses and overall increased roughness of a printed part. The addition of nanoparticles can minimize these temperature-based variations due to smaller CTE of the feedstock material and better thermal transport. Improving surface finish both for FDM and SLS due to nanofillers such as CNTs, graphene, layered silicates and other anisotropic nano-constituents is accomplished via formation of thin (10s of nanometers) sheaths with reduced surface roughness. The smoother and more uniform surface of an extrudate from the nozzle enables higher precision printing of roads with minimized defects.[95] Printing accuracy was also improved via the addition of gamma-$Al_2O_3$ nanowires to a photopolymer (reduced distortion of printed flexible parts). Printing of the composite inks via photopolymerization with 12wt% of the nanofiller also leads to 50% improvement in the mechanical properties over the unfilled system.[96] On the other hand, nanoparticles can also affect accuracy negatively as was shown in an SLA process in which layered silicate and attapulgite nanoparticles were added to the resin. The reason is attributed to incomplete curing in the presence of the scattering and shielding nanoparticles that absorb the incoming light and prevent curing throughout the printed roads.[63]

### 2.1.6 Printing submillimeter features

Although this review focuses on additive manufacturing of larger parts, there are important sub-millimeter techniques that employ nanoparticles, and have potential to be transformed to larger scales (mm, cm) in the future. One example is reduced graphene oxide (GO) nanowires that are additively manufactured via a meniscus-guided growth using micropipette solution of graphene with subsequent reduction of GO by thermal or chemical treatment to create stretchable interconnects for electronic devices and transducer for gas sensors.[97] In another study, highly conductive gold micro wires and walls were printed via an inkjet technique with gold nanoparticles of 2-4nm in size. The gold nanoparticles were dispersed in an organic solvent with a weight fraction of about 30% (~2vol%) and were annealed by a laser treatment [98] to form wires seen in **Figure 11**.

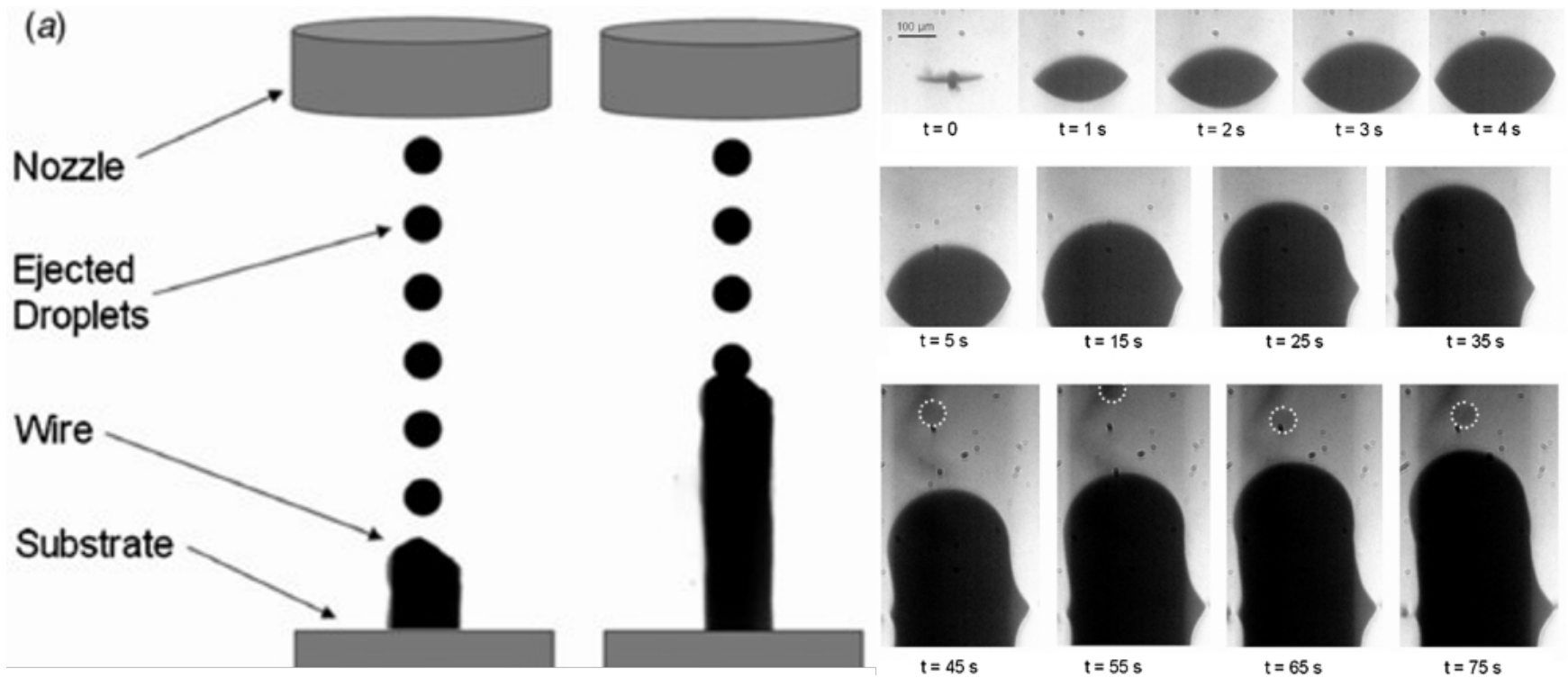

**Figure 11**: Schematic showing the growth of micro-wires via inkjet deposition. Optical images of Au micro-wire printing at room temperature as function of time.[98]

Gold nanoparticles were formed in situ during multiphoton lithography using gold salt containing crosslinkable triacrylate solutions to build three-dimensional structures on the micron scale.[99] Conductive 3D structures were formed via UV polymerization of emulsion-based inks with silver nanoparticles at low temperatures that have been inkjet deposited. After rapid polymerization of the monomers, the structure is developed and dipped into a salt solution to induce the sintering process, while retaining the shape.[100]

The use of nanoparticles in inkjet printing for biosensor applications has been thoroughly summarized in a critical review article, including CuO, Ag, Au, dye-encapsulated polymer, $TiO_2$ and silica nanoparticles. In most of these cases, the function of the nanoparticle is to provide conductivity.[101] The use of inkjet printing enables users with a straightforward and inexpensive alternative that has a low risk for contamination, minimal waste, the ability to use multiple sensing nanoparticles, and the capacity to enable precise spatial control and gradient creation, which can be limited in other AM methods.

In another project, nanofibers are additively build up into micron sized walls via electrospinning of polymer solutions onto one-dimensional conductive paths even without the use of a moving head.[102] The combination of two versatile techniques, namely direct-write maskless UV patterning and layer-by-layer assembly (LBL), were used to create MEMS nanocomposite thin films. Single-walled carbon nanotube (SWCNT) and gold nanoparticle LBL nanocomposites were assembled with chitosan and made into patterns used in flexible antennas and neuroprosthetic devices.[103] Thiele et. al report on producing 12μm thin fuel cell membranes by printing Nafion ionomer dispersions onto mats of nanocomposites made of Poly(vinylidene fluoride-co-hexafluoropropylene) nanofibers.[104] In this case, the nanoelements are not directly used in the printing process, rather 3D printing is used to accomplish an otherwise difficult procedure.

Nanotribological printing (NTP) is a nanoscale process that was discovered by Khare et al. by using standard contact-mode atomic force microscope (AFM) for nanoscale printing.[105] Nanoscale printing is achieved due to the normal and shear forces applied in the NTP process. The tribomechanical and tribochemical surface interactions at the contact between the substrate and AFM probe forms the designed pattern. The "ink" that is used consists of a carrier fluid and nanoparticles which surround the AFM probe. In future research works, this novel manufacturing process has the capacity to utilize multiple nanomaterials at once through subsequent patterning of the materials on a variety of substrates.

### 2.1.7 Nozzle & Print Path Induced Alignment

Using shear forces to align nanoparticles has been effective in tailoring the finished nanocomposite's properties without using external forces, such as magnetism or electrical currents. This technique is best utilized with extrusion and direct write based additive manufacturing methods since it makes use of the nozzle to manipulate the shear forces in the nanocomposite material. The relatively high shear and extensional flow, that can develop during the additive process (e.g. in the printing nozzle), is a rather ubiquitous and intuitive way to orient high aspect ratio nanoparticles present within the "printing" medium. This synergistic strategy for additive manufacturing with nanocomposites bridges the areas of process control and morphological control through use of various nozzle geometries and shear forces.

Nanomaterial inclusions in FDM printing can contribute to improved flow properties of polymers that have high viscosity, such as PEEK. Golbang et al. conducted a study to observe the effects of Inorganic Fullerene Tungsten Sulfide (IF-$WS_2$) nanoparticles that were melt compounded with PEEK and extruded into filament for FDM printing.[106] Without IF-$WS_2$, PEEK has a low shear rate which causes back pressure on the filament during printing and results in the filament buckling or having reduced flow. At a 2 wt% loading, the IF-$WS_2$ particles were shown to increase the melt viscosity 25% and acted as a lubricant for the nanocomposite when it was extruded through the FDM nozzle. Fewer voids were visible in printed samples with 1 and 2 wt% IF-$WS_2$, compared to neat PEEK, signaling higher quality prints were achieved. Additionally, the nanoparticle addition did not cause a significant increase in melting temperature or crystallization, which is beneficial to the FDM process and printed parts.

Compton et al. showed that SiC nanowhiskers (0.65 μm average diameter; 12 μm average length) and short carbon fibers (CF) (10 μm average diameter; 220 μm average length) align within a micro-nozzle and along the printing direction, during printing of an epoxy-based "ink" formulation.[47] Consequently, anisotropy in mechanical properties (ratio of Young's modulus parallel and perpendicular to reinforcement direction) was achieved, increasing with reducing nozzle diameter and with the addition of carbon fibers as secondary filler. A maximum Young's modulus, longitudinally to the printing direction, of 24.5 GPa was achieved for 31 vol.% of solid content (28 vol% SiC + 3 vol% CF), which is equivalent to a 9 time increase compared with casted samples composed of the pure epoxy resin. This increase is far from the maximum mechanical reinforcing efficiency which can be predicted from micromechanical. Despite this enhancement, defects, like porosity, were still present in the specimens that caused lower tensile strength (96.6 MPa for samples containing 28 vol% SiC, 66.2 MPa samples containing 28 vol% SiC + 3 vol% SiC) than the value of pure epoxy resin (71 MPa).

Andersson et. al used a top-down and bottom-up approach to form a polymer ceramic composite by alignment of self-assembled polymerizable lyotropic liquid crystals via 3D printing. Additional reinforcement of the polymer matrix is done via in situ mineralization of bone-like apatite.[107] The authors create a proof-of-concept by making large scale composites that resemble heterogeneous tissues like bone-cartilage interfaces. Microcracking and toughening mechanisms within the printed specimen were also studied and point toward a combination of molecular + nanoscale structure (intrinsic) and micro + macroscopic architecture (extrinsic) contributing to the crack resist character of the composites.

The utility of nanoparticle shear alignment is not only restricted to traditional thermoplastic and ceramic materials but can be applied to hydrogels with nanofillers as well. In the work by Zhao et al., CNT is embedded in 2-hydroxyethyl methacrylate (HEMA)-based gelation system and manufactured in an extrusion-based process.[108] The process uses two independent syringe pumps, one with the HEMA-CNT slurry and the second with APS initiator solution. The material is fed into a Delta 3D printing platform, which consists of three parallel arms that are attached to a platform, like an FDM style device.  This technology allows the nanofillers to undergo shear induced alignment in the nozzle and retain this quality in the finished printed products. The directed CNT alignment is shown to alter the ultimate strength and exhibits strain-rate-dependent hardening. Furthermore, the orientation of the aligned CNTs has a large influence on the hysteresis under dynamic loading and allows the opportunity for aligned hydrogels to be used in tissue engineering or shape memory composites.

Due to their notorious difficulty to 3D print, hydrogels must incorporate nanomaterials to aid in the manufacturing process. In Jin et al.'s study, a type of Laponite nanoclay and graphene oxide (GO) were used as a reinforcing material in the NIPAAm hydrogel matrix.[109] The use of these nanomaterials is not limited to this matrix alone, as seen in previous sections, but also with other hydrogel matrices that would otherwise be difficult to print. The printing process utilizes two dispensing pumps to print alternating filaments of NIPAAm-Laponite and NIPAAm-Laponite-GO nanocomposites, each of which is subjected to shear-induced alignment along the printing direction (**Figure 12**). The self-supporting property of the Laponite nanoclay makes printing possible as the two hydrogel materials remain liquid and do not undergo any curing. After printing, the structure is exposed to UV radiation for final solidification. This method of fabrication opens the door to hydrogel nanocomposites with multifunctionality and stimuli-responsive properties in future endeavors.

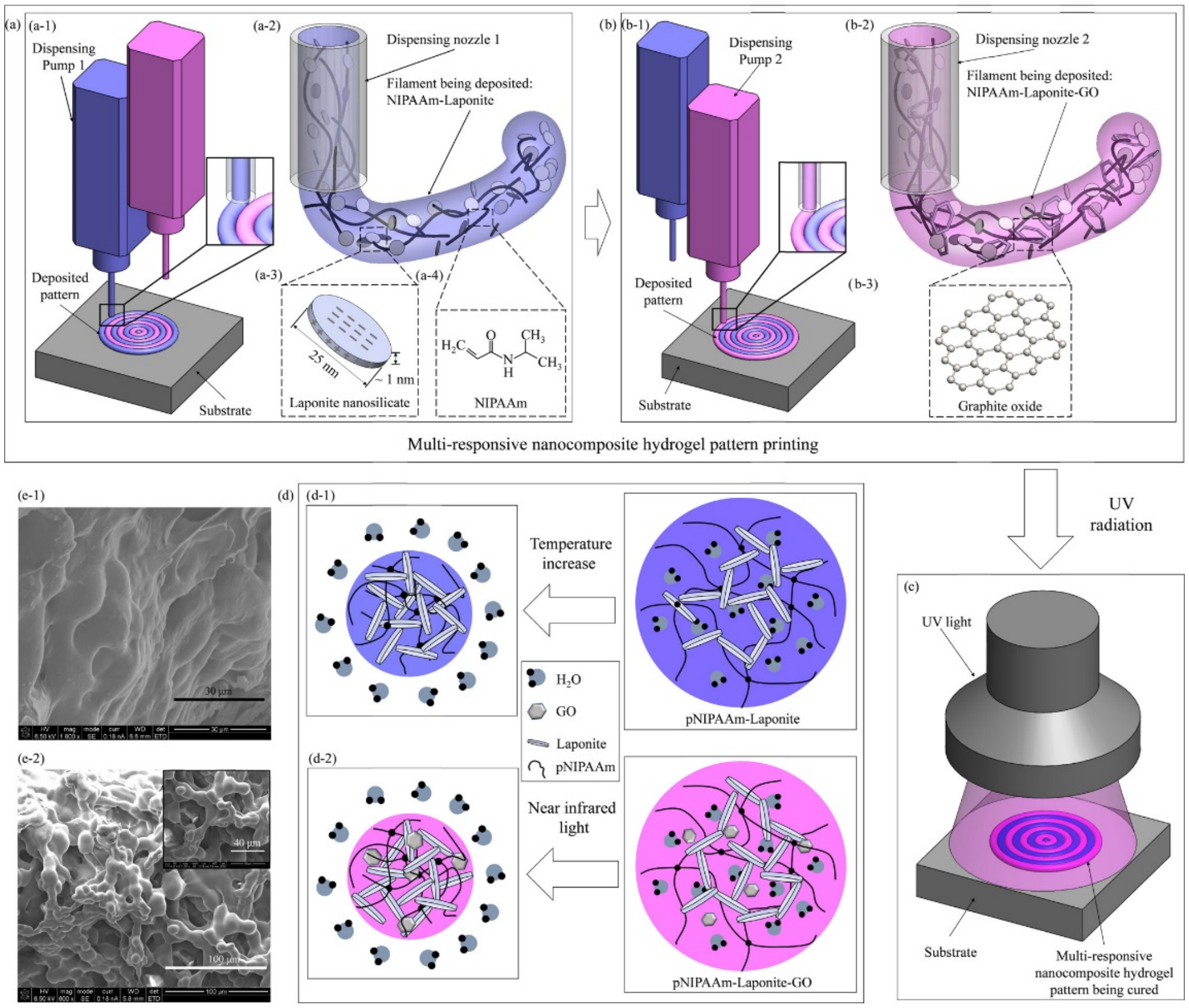

**Figure 12**. Schematic of printing nanoclay-based pNIPAAm nanocomposite hydrogels. (a) Printing process of NIPAAm–Laponite nanocomposite hydrogel precursor. (b) Printing process of NIPAAm–Laponite–GO nanocomposite hydrogel precursor. (c) UV curing of deposited structures. (d) Mechanism of nanocomposite hydrogel deformation. (e-1) SEM image of pNIPAAm–Laponite nanocomposite hydrogel (scale bar: 30 µm). (e-2) SEM images of pNIPAAm–Laponite–GO nanocomposite hydrogel (scale bars: 100 and 40 µm (inset)).[109]

A similar flow induced orientation, as well as its dependence on nozzle size and printing speed, was also observed by Gladman et al. for a soft acrylamide hydrogel filled with nanofibrillated nanocellulose (NFC).[110] The degree of alignment of NFC along prescribed printing directions, was demonstrated to control anisotropy in both mechanical (stiffness) and swelling behavior (4 times difference in swelling strain between transverse and longitudinal directions).In these studies, tuning the viscoelastic properties of the printing “ink” (i.e. high shear thinning and shear yield strength), was identified as an essential processing and was achieved by using rheological additives, typically nanoclays, fumed silica or high molecular weight polymers.[47, 62, 111]

The effect of nozzle shape on the alignment of Silver nanowires (AgNWs) in a silicone rubber matrix  and its effect on dielectric properties were explored by Kim et al.[112] The nanocomposites were extruded through two nozzles, one with a circular cross section and one with a flat or rectangular cross section. The morphology of the samples shows that the circular nozzle aligns the AgNWs more effectively resulting in higher dielectric permittivity compared to that of flat or rectangular nozzle where the nanowires are randomly aligned.

The effects of nozzle geometries and nanofiber concentrations were also studied by Papon et al. using PLA and carbon nanofiber (CNF) composites for FDM.[113] Contrary to Kim et al.'s, a square nozzle geometry was shown to improve modulus and reduce voids compared to a circular nozzle. A 12% enhancement of the material's modulus at the 1% CNF loading is observed compared to neat PLA. The reduced voids that are observed can be attributed to the square shaped road that results from extrusion from the square nozzle. The approximated inter-bead void areas within the printed samples accounted for 2% of the total area. Although there are differences in the printed parts due to nozzle geometry, the CNF are aligned within the PLA matrix using both nozzles.

Nozzle dimensions are not only limited to FDM style printing but can be altered in direct write printing processes that use needles. Single filaments of acrylated epoxied soybean oil (AESO)/polyethylene glycol diacrylate (PEGDA)/ nanohydroxyapatite(nHA)-based ink for bone defect repairs were studied with various needle diameters, ranging from 0.21 mm to 0.84 mm, to evaluate the effect on mechanical properties of the printed parts.[114] It was discovered that the tensile strength was increased as the diameter of the needle was decreased. The strength for the 0.84 mm diameter needle was 26.3 MPa compared to 48.9 MPa with the 0.21mm diameter nozzle. This dramatic improvement in tensile strength is attributed to increased crystallinity and fewer defects in the matrix material.

In traditional casting applications with nanocomposites, nanofillers are randomly aligned and do not exhibit tailored multifunctionality that is possible with these types of materials. To govern the alignment of the nanofillers, spatially controlled alignment of piezoelectric nanowires was accomplished in a study by Malakooti et al., in which shear forces in the nozzle of a direct write printer allowed $BaTiO_3$ NWs to self-align during extrusion.[115] The alignment of the barium titanate nanowires had a strong bearing on the electromechanical properties of printed parts when arranged using print angles. The alignment of the nanowires in the longitudinal direction increased the maximum generated AC power more than 700%, compared to solvent cast samples with random alignment.

As stated previously, 3D printing is not constrained only to layer-by-layer processes while exploiting the shear-thinning properties of nanocomposites. Using additive manufacturing techniques, a 2-dimensional material, such as a nanosheets of boron nitride (BN) or graphene, can be assembled into macroscale structures with functionality.[116] BN nanosheets were vertically aligned through shear thinning behavior. This behavior allowed a fine nozzle to orient the BN nanosheets in a vertical direction. The vertically aligned structures that were produced were able to conduct through plane heat and dissipate it from the heat source. This method of manufacturing can be used in electrical and electronic systems and where anisotropic structures are desired. Similar effects were observed in a study by Haney et al. in which the authors used direct ink writing of 7-18 wt% loading of graphene in Epon 862 to create coupons with 10 times improved electrical conductivity ($10^{-3}$ S/cm) due to a shift in the orientation of the graphene fillers as a function of print speed (5-40 mm/s), with respect to printing direction.[117] Siqueira et al. reported use of cellulose nanocrystals (CNC) as a nanofiller in 3D printing "ink that did not require any rheological additives.[118] In this case, aqueous suspensions with relatively high nanofiller loadings, with or without soluble polymers, were directly used as "ink", as higher filler contents were achievable due to liquid crystalline behavior of CNC. Above a critical concentration (~ 5 wt% for a CNC with aspect ratio of 18), CNC ordered in anisotropic crystalline regimes in aqueous suspensions, as confirmed by birefringence tests. The ordering of CNC particles was further enhanced under flow, inducing a higher nanoparticle orientation along the printing direction, compared to solution-cast films. confirmed by optical birefringence, AFM and 2D WAXS. The CNC alignment was further affected by the printing of successive layers. The interface between two filaments intermixed during direct writing, which on the one hand, promoted adhesion but, on the other hand, misaligned CNCs.  While not stated by the authors, it is possible that misaligned CNCs may promote adhesion via mechanical interlocking. Addition of 10 wt% CNC increased the Young's modulus along the longitudinal (printing) direction increased by 30% and ~1000% by for stiff (~0.9 GPa) and soft (4 MPa) polymer, respectively. The Young's moduli in the transverse direction did not increase more than 20%, which was explained by the CNC misalignment at the interface between two printed filaments.

In a combination of print path and nanoparticle volume fraction, AlMangour et al. have shown that they can tailor densification level, solidification microstructure, crystallographic texture, and anisotropy of mechanical properties of fabricated parts of sintered TiC/316L stainless steel nanocomposites.[119] Ductility and strength were improved in the SLM processing of novel composite powders with uniform $TiB_2$ nanoparticle distributions, in addition to much better laser absorptivity and therefore improved SLM processing.[120]

### 2.1.8 Real Time Monitoring for Process Control

As important as the innovation in methodology and materials is in this field, the progression of monitoring processes and developing real time measurements for process control with feedback, is equally important. While the more basic in-situ measurements such as temperature and dimensional control are already implemented in production type machines, this chapter focuses on the advanced tool development for feedback control. While we believe that in-situ monitoring does not truly fall under the scope of this review and necessitates a separate standalone review, we added a section with focus on methods that explore the nature of this Process Control section and the following Morphology Control section. In a study by Johnson et al., thermoset nanocomposites were monitored using X-Ray photon correlation spectroscopy (XPCS).[121] This technique explores the effects that print and substrate parameters have on the morphology and dynamics used in the direct write process of thermosetting nanocomposite or conductive inks [122]. During experiments, filaments were deposited across the X-ray beam path, with the print head moving in an x-translation direction (**Figure 13**). Once the filament has been deposited, there is no movement outside of the viscoelastic drag of the material that is an unavoidable part of the printing process. By evaluating the dynamics of the nanocomposite as a function of both space and time, the material's temporal evolution and anisotropy of the dynamic recovery can be characterized. The dynamics of the printed material have been shown to strongly depend on the flow profile from the print head during extrusion as well as the velocity and shear stresses experienced in the nozzle. However, the alignment of the layered silicate particles as well as their recovery dynamics have a greater impact than just the extrusion itself. The alignment of the layered silicate is due to the small radius of curvature where the material is being deposited on the build plate and experiencing viscoelastic drag by the downstream side of the nozzle. The XPCS monitoring process can be used to precisely tailor properties of the nanocomposite and the optimization of thermoset materials and nozzle design.

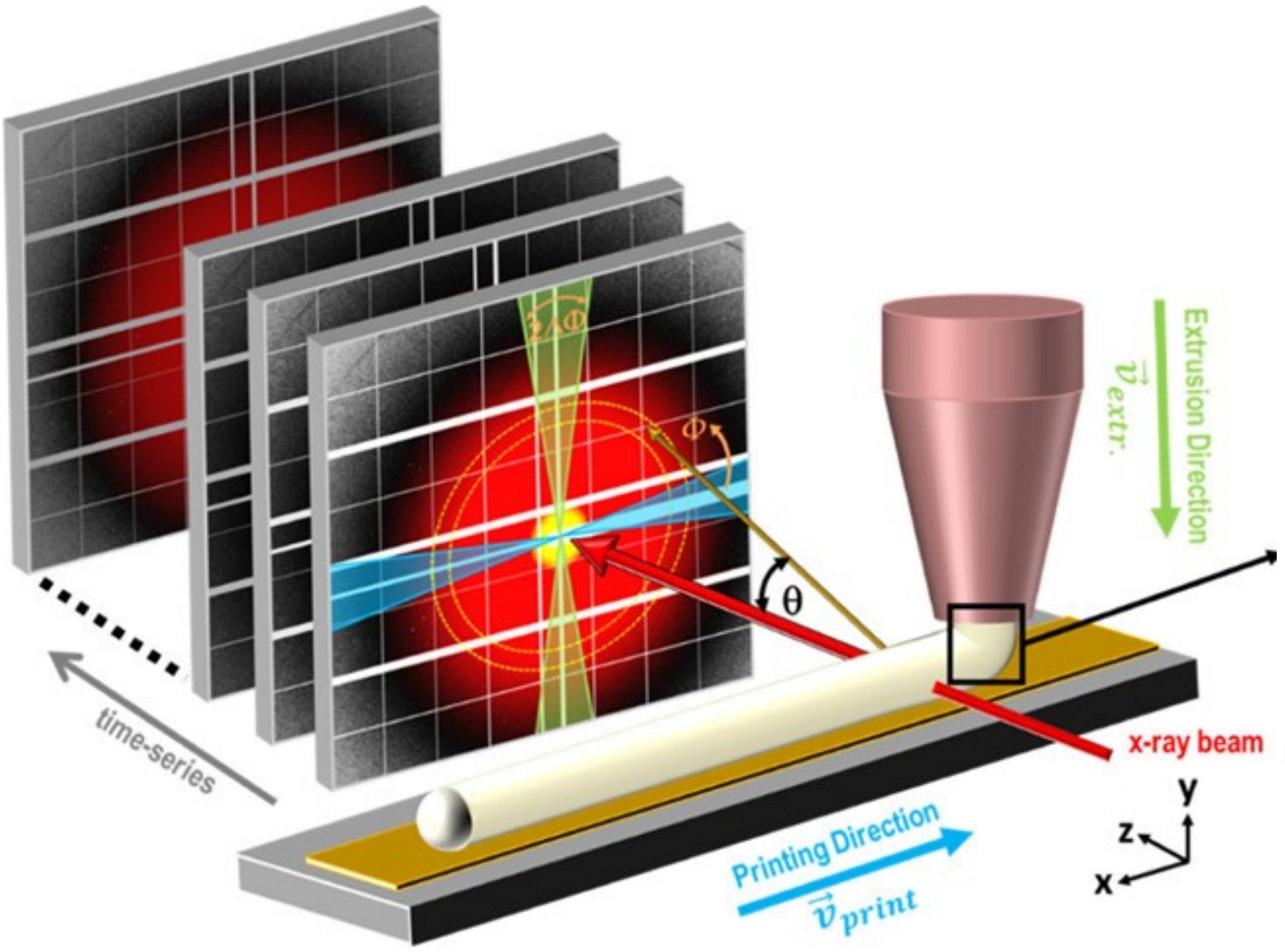

**Figure 13**: Schematic of the extrusion printing process and the scattering geometry for XPCS in SAXS geometry. The material exits the nozzle at the same speed as the nozzle is moving with respect to the build plate, |v→extr.| = |v→print| = 0.1 mm/s. The print (ϕ = π) and extrusion (ϕ = π/2) directions are normal to one another, and the print head and build plate are placed to pass orthogonal to the incoming X-ray beam. The scattered X-ray beam corresponding to a certain q vector, defined by the scattering angle θ, is shown in orange, and the corresponding area q ± Δq on the scattering image is marked with dashed orange lines. The blue and green highlighted sections correspond to the Φ-directions parallel to the extrusion and printing directions, respectively, with widths 2ΔΦ. White lines on the scattering image indicate masked pixels (beamstop, detector module, and chip gaps) that were not used in the analysis.[121]

Similarly, Raman spectroscopy and X-ray microbeam SAXS analysis were used to evaluate FDM printed polypropylene/graphene nanocomposites.[123] It was demonstrated through rheological measurements that the presence of graphene induced shear thinning during extrusion at concentrations of 10% nanofiller or greater. Additionally, the graphene was oriented in the direction of extrusion from the shear forces that the material experiences. The graphene orientation in the filaments provided a 267% increase in conductivity at the 10% loading concentration. The Raman spectroscopy exhibited that the extrusion process exfoliated the graphene, which effectively increased the attractive interactions between the nanomaterial and the matrix and increased the viscosity of the material.

In operando phase contrast imaging (radiography) was used in the monitoring of fiber reinforced epoxy inks. [124] While this work focused on larger scale silicon carbide and carbon fibers (5-7 um diameter), this measurement tool is also applicable to nanocomposites to track the mortality of defects such as voids and agglomerates. The detailed understanding of nucleation, evolution and final state of voids is critical to reduce failure sites in the final part and improve its performance. This measurement tool can be coupled to fast real-time evaluation of nozzle geometries coupled to printability of the feedstock used during the AM process. This technique provides an instant map of velocity and alignment within the nozzle, typically within a 3x3mm window.

High resolution micron scale scanning and mapping of morphology in printed parts is equally important to understand interfaces between roads and layers. Micro-beam X-ray scanning was used by Trigg et al. to show the nanoparticle alignment in a 3D printed composite ink with a layered silicate as rheology modifier.[125] This technique revealed the complexity of morphology which evolves due to alignment of layered silicate nanoparticles within the nozzle, and during the deposition and the interaction of the nozzle with already printed material. A map of Angstrom to nanometer scale morphology at a resolution of 5 micrometer is completed via automated scanning and data analysis using machine learning methods.

### 2.1.9 Postprocessing

While many of the more conventional additive manufacturing processes, such as FDM or SLS only require removal of support material and surface finish procedures, a post-print cure process has to be done for thermosetting resins.[126] For the manufactured parts to be exposed under uniform temperatures during a post-cure cycle, improved energy transfer from the heated environment to the resin, especially into thicker sections, is advantageous. Inclusion of particulates such as reinforcement fillers (chopped carbon fibers, SiC whiskers) and nanoparticles (carbon nanotubes, nanofibers, metal oxide and metal and polymer grafted nanoparticles) enable better control during this post-printing step, either by improved heat transfer (conductivity), by reducing cure shrinkage and spring-back due to inhomogeneous temperature gradients throughout the additively manufactured parts, or by limited physical aging effects that are enhanced by the AM process (micro-cracking, embrittlement, delamination).[47, 127, 128] While the need for better post-processing control using nanoparticles has been recognized in the AM community, very little has been published on specific systems that demonstrate this approach successfully and most of this is notional speculation at the time of writing.

In summary, manipulating the process itself through means of shear thinning, energy transfer, reaction kinetics, submillimeter features, printing precursors, post processing and nozzle induced alignment, has been proven in a multitude of studies to alter the behavior of the nanoparticles, matrix material, or the entire nanocomposite itself during manufacturing. With this is mind, there is room for improvements in existing processes as well as development of new processes that could further improve the relationship between the nanocomposites themselves and the AM method being used. While there is a rich range of processing parameters that can be positively affected by the inclusion of nanoparticles as shown in this section, the presence of these nanoparticles in the final product also opens opportunities for morphology control within roads and layers, and at interfaces. Structure and morphology control are used for numerous reasons that depend on the end-use application of an additively manufactured part. In many cases, it simply addresses mechanical properties of the resulting parts. While volume fraction is the most basic parameter that determines the mechanical properties of printed parts, the orientation or alignment of nanoparticles into strings or sheets or the inherent shape anisotropy of the particles plays a major role in true control of micro-mechanical properties when coupled with tool-path optimization in an additive manufacturing process. The next chapter discusses some important work in this area.

## 2.2 Morphology Control

While the morphology within additively manufactured parts is primarily controlled via rheology and flow-induced fields as described before, there are additional methods that can be applied to alter the morphology on both nano and micro-scale using external fields or templating methods. External fields allow the freedom to create highly tailorable materials with custom and complex geometries. These are methods that have been applied in conventional extrusion, jetting, spinning, and dispensing type applications. The local nano and micro scale directionality of functional and reinforcement filler can be modified beyond the existing shear flow, which has limitations based on material properties and print speed. In some cases, the primary particles might be inert/non-responsive to external fields but can be modified and functionalized with responsive material to enable that additional control. In other cases, an additional inert material may guide or template the orientation or location of primary particles.[129] Though the strategies used in this chapter can significantly improve the relationship and outcomes of AM and Nano, there is still opportunity for exploration in this area, which is shown by the fewer published works available.

### 2.2.1 In-line Directed via External Field

In the case of most nanofillers used with field-assisted processes, the nanoparticles themselves are responsive to either magnetic or electric forces being applied or the addition or removal of heat. The use of magnetic and electric fields is the most commonly used in contemporary efforts to align the particles in a specific orientation and can be retrofitted into existing AM practices with relative ease, as with changing the environmental temperature of the operation. Though there are other external fields that can be applied, such as acoustic, they are not as easy to implement. Additionally, mechanical methods can be used to induce nanofiller alignment, however it is seen largely through means of nozzle induced alignment, as discussed in the previous section.

One strategy to bypass the requirements on the viscoelastic properties of the printing "ink", is to freeze it as it is ejected from the nozzle. Zhang et al. developed a modified 3D printing technique, with a cold sink (-25 °C), onto which an aqueous "ink", containing graphene oxide (GO), could be inkjet printed into 3D structures, even though it had low viscosity and Newtonian behavior.[130] However, the solidification of the aqueous "ink" provided an additional benefit. Highly ordered and anisotropic assemblies of GO platelets were formed and templated at the crystal boundaries as ice was formed during freeze casting. After sublimation of ice by freeze drying, porous honeycomb structures with oriented GO were created. Moreover, because of the possibility of partial melting before sublimation, combined with the low viscosity of the melting liquid, the void level and the bonding of different layers could be controlled to the point of eliminating inter-layer boundaries. The boundary elimination was demonstrated by studies of the interface microstructure and compression tests. Partial fracture of porous 3D printed structures occurred when compression strains exceeded 50%, and fracture cracks occurred along the shear propagation direction rather than along inter-layers. Interestingly, freeze-casted 3D printed GO structures showed a higher specific stiffness than the bulk counterparts due to the orientation of the GO nanoplatelets.

Another method to control both three-dimensional orientation and distribution of the reinforcing particles during AM is to use an external magnetic field. Unfortunately, traditional reinforcing particles are often diamagnetic, requiring high magnetic fields (~1 T) for alignment.[131] Erb et al. have shown that coating with superparamagnetic (iron oxide) nanoparticles can overcome this limitation.[132] Both theoretical and experimental work demonstrated that, for certain ranges of reinforcing particle size and iron oxide nanoparticle surface coverage, very low magnetic fields (~0.8 mT) were sufficient to align the reinforcing particles, orders of magnitude lower than the magnetic field of rear-earth magnets (~200 mT) and common solenoids (~20 mT). Platelets and rods of few microns in length (5-10 μm), and only 0.5 vol.% iron oxide surface coating, were found to require the minimum magnetic field for orientation, order of magnitudes lower than smaller or larger particles, which were instead dominated, respectively, by thermal and gravitational energies. This effect, defined as ultra-high magnetic response (UHMR), was then utilized to prepare AM composites based on different polymer matrices (polyurethanes, polyacrylates, polyvinyl alcohol and epoxies) with different orientations and distributions of reinforcing particles. Uniform, linear static and rotating magnetic fields as well as spatial magnetic gradients were used to create various structures (**Figure 14**). By controlling the orientation and position of the reinforcing element, physical properties including stiffness, strength, hardness, wear resistance, and the shape memory effect could be tuned. The increase of the yield strength of the anisotropic composites could be interpreted with a simple rule-of-mixture analytical model, assuming pull-out mode of fracture and an interfacial shear strength lower than the one expected for the pure polymer. This suggests a weak polymer/reinforcing particle interface or the presence of other defects (e.g., porosity).

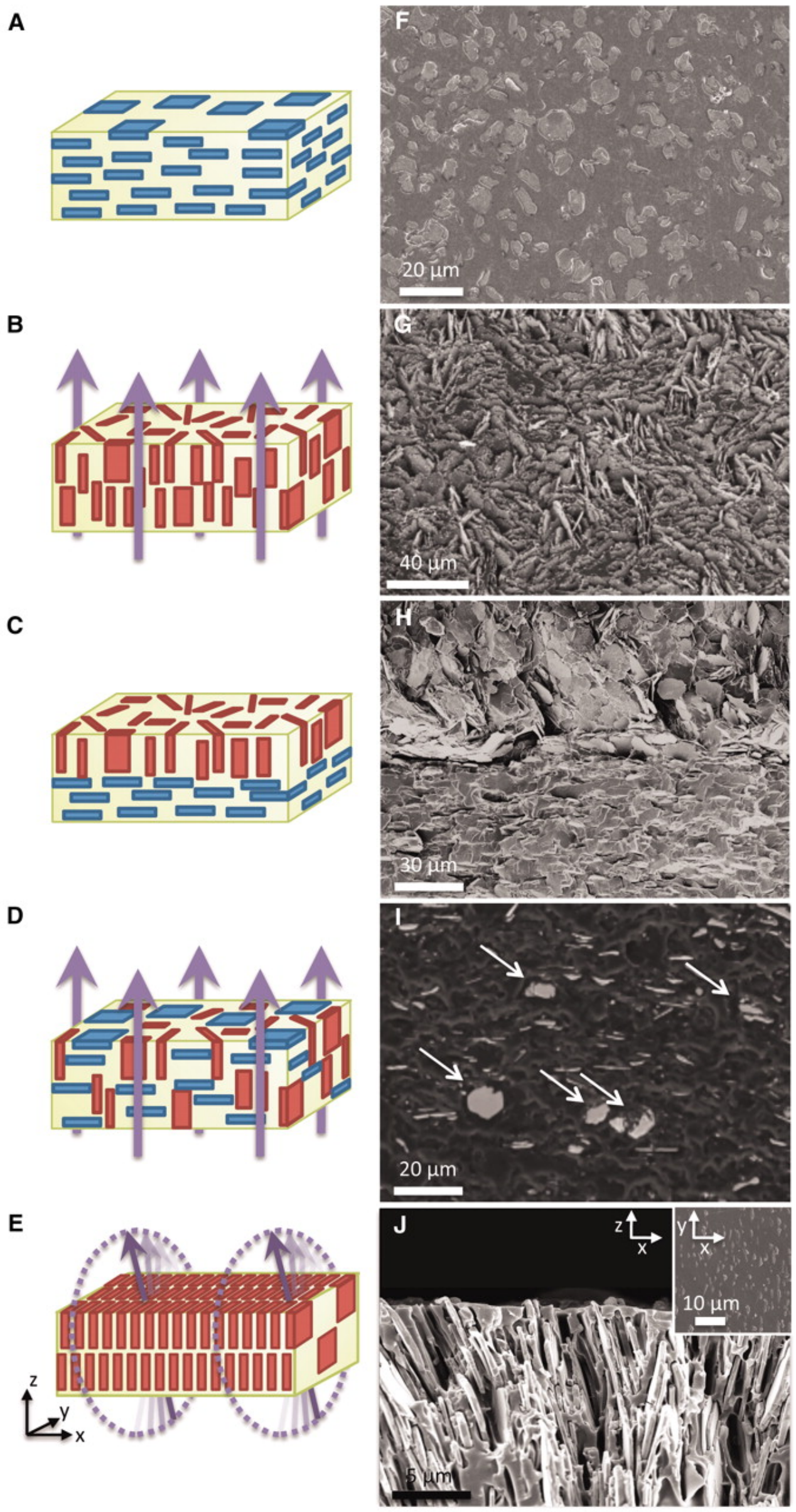

**Figure 14**: In-plane and out-of plane alignment of UHMR alumina platelets in polyurethane-based composites. (A and B) Schematic and (F and G) top-view scanning electron micrographs (SEMs) of in-plane and out-of-plane reinforced composites (20 vol % Al2O3 in polyurethane), made without and with an out-of-plane magnetic field, respectively. (C) Schematic and (H) SEM of cross section of laminated layers of in- and out-of-plane reinforced composites (20 vol% Al2O3 in polyurethane). (D) Schematic and (I) SEM of cross section of mixed alignment with 5 vol % highly magnetized (1 wt % Fe3O4/Al2O3) and 5 vol % weakly magnetized (0.1 wt % Fe3O4/Al2O3) alumina platelets in polyurethane produced with sequential magnetic field applications. Arrows indicate platelets in the second orientation direction. (E) Schematic and (J) SEM of cross section of an alumina-polyurethane composite

formed under rotating magnetic field that allows for ultrahigh packing fractions. Inset shows top view of composite.[132]

Unfortunately, the UHMR above is only observed for a narrow reinforcing particle size range, for relatively large particles (~1-10 μm) or low aspect ratio (~10-40).

Tognato et al. use a low-intensity magnetic field along with the self-assembly capability of iron oxide nanoparticles (IOPs) to create oriented filaments with aligned mosaic IOPs within a gelatin methacryoyl matrix.[133] Once self-alignment was achieved, the temperature was decreased below the melting temperature to preserve the assembled filament structures before being exposed to UV light to finalize cross-linking. It was suggested that this novel processing technology could be used in bioinspired soft robotic systems. These systems have the functionality from the nanocomposite to create tissue scaffolds that do not rely on external biochemical effectors.

Pan et al. have used magnetic field assisted projection stereolithography to direct nanoparticles for 3D printing of smart structures.[129] The external magnetic field enables the controlled distribution of magnetic nanoparticles with various patterns. A mask image was then projected to cure the photopolymer, and the developed process is capable of achieving various magnetic particle-filling rate, filling pattern and structure.[134]

A more versatile method was recently reported by Yang et al., demonstrating an electric field assisted AM method to accurately control the orientation of high aspect ratio nanoparticles, specifically, multiwalled carbon nanotubes (MWCNT).[135] Depending on the configuration of electrodes, parallel alignments, radial alignments, and circumferential alignments were successfully achieved using, respectively, parallel plate electrodes, needle-arc electrodes and two needle electrodes (**Figure 15.a**), by balancing three key forces: torque, coulombic and electrophoresis forces. Depending on the orientation imposed, the mechanical properties could be tuned (**Figure 15.b-c**).

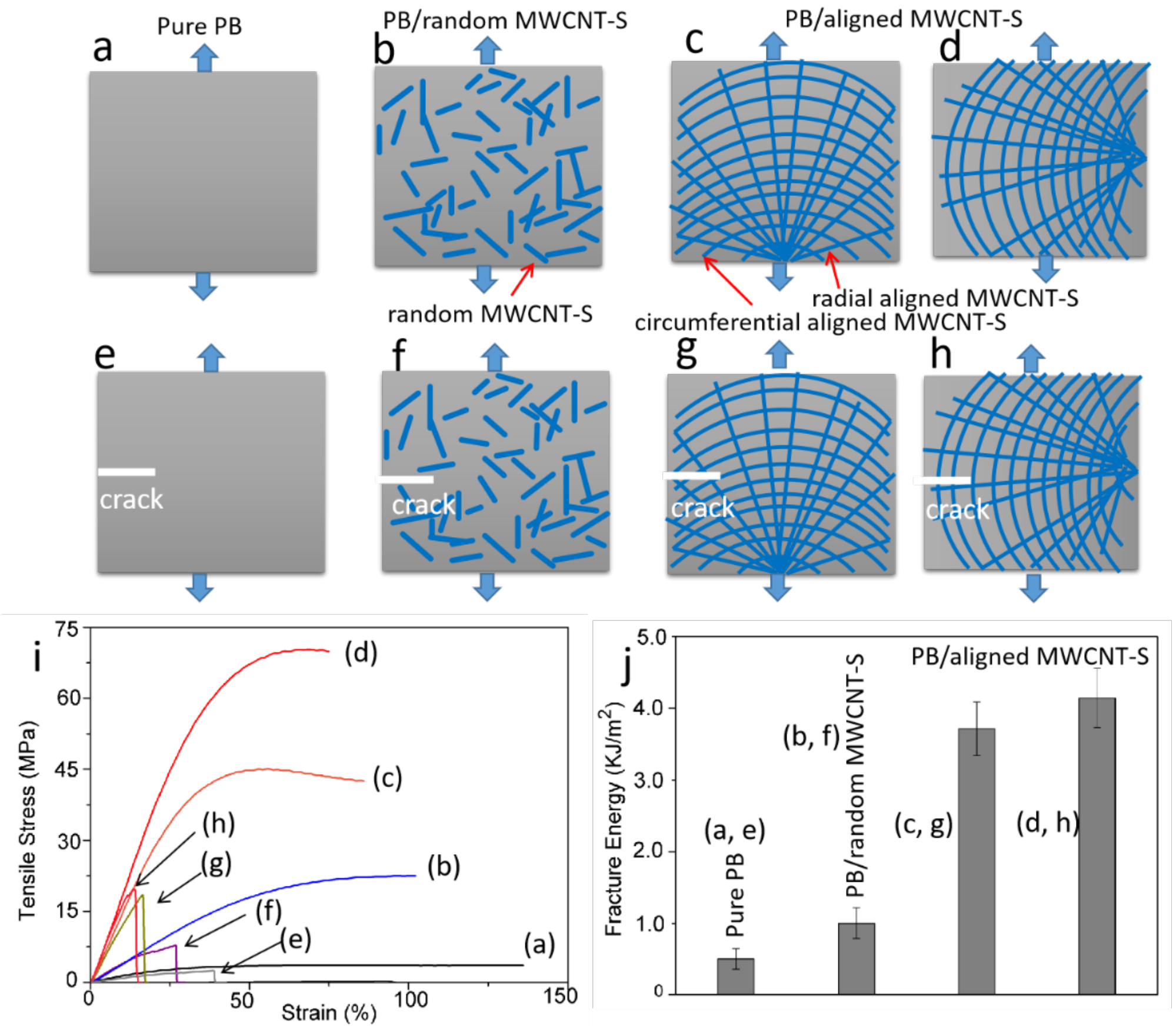

**Figure 15**: Schematic diagram of test samples for fracture energy. a) pure PB and with a pre-cut crack e), b) PB/random MWCNT-S and with a crack f); c) PB/aligned MWCNT-S and with a crack g), tensile force parallel to the radial alignment; d) PB/aligned MWCNT-S and with a crack h), tensile force parallel to the circumferential alignment; i) stress-strain curves of these samples; j) comparison of fracture energy. [135]

Begley et. al have shown in-line controlled microstructures via acoustic focusing of two-phase materials during direct ink writing. Acoustic waves are used to focus nanoparticles within the printed roads during printing.[136] Increased silica content in the direct write ink increased the viscosity as well, which improved the preservation and integrity of printed shapes, and lays the foundation for future works using acoustic waves for field-assisted control.

Electric current is used by Bodkhe et al. to create piezoelectric sensors with PVDF and $BaTiO_3$ nanoparticles.[137] A custom printing apparatus with a robotic arm and dispensing system was enabled with a DC power source to enable in-line high-voltage poling by applying the electric field between the printing nozzle and metallic tape attached to the print bed. The ink developed in this study allows for producing self-supporting 3D scaffolds with some degree of geometric complexity. A 10 wt% $BaTiO_3$ NPs and an electric field of $1MVm^{-1}$ showed 300% improvement in piezoelectric charge output, compared to neat PVDF. Through this synergistic use of additive manufacturing with PVDF-$BaTiO_3$ and simultaneous poling, a close control on porosity and an even dispersion and distribution of nanofillers for multifunctional nanocomposite sensors was obtained.

In an innovative way to circumvent this generic 2.5D printing problem, Shariatnia et al. used an integrated sprayer within a FDM modeling 3D printer to deposit cellulose nanocrystals (CNCs) on the surfaces of printed thermoplastic filaments.[138] Prior to this study, there had not been much exploration in harnessing the use of a spray system for a nanoparticle suspension in conjunction with an FDM printer. The microstructure study showed that CNCs act as nanopins between adjacent printed layers and enhance the interlayer adhesion and overall strength of printed parts by 40%. This new technology enables the use of spraying nanoparticles beyond just CNC, so long as they are suspended in an aqueous solution, to be utilized for improved part quality and functionality.

Within this section, we have highlighted the main research efforts in this area with a specific focus on external field assisted morphology control of AM structures. A separate and upcoming research field is constituted by stimuli-responsive nanocomposites that utilize external forces including fluid shear, evaporative, acoustic, electrical, magnetic, optical, and thermal patterning review. The readers are referred to the review paper by Elder et al.[139]

### 2.2.2 Directed via Post-treatment

Post-treatment of printed material beyond simple finishing and machining, such as annealing or external field aided annealing (electric, magnetic, acoustic, and optical) may enhance the morphology or even induce the formation of structure within printed parts. An example is that the inclusion of polymer-grafted nanoparticles in additively manufactured parts in an ordered fashion will enable control of physical aging due to confinement effects of the polymer matrix within discrete interstitial particle spaces. This will have effects on fit, form and stability of printed parts and can be used in combination with print path topology to optimize spring-back, warpage and delamination issues.[140]

Through the assistance of microwave heating, an accelerated annealing strategy is used with the goal of enhancing crystallinity and mechanical properties of PLA-cellulose nanocomposites fabricated using FDM.[141] Carbonized cellulose nanofibers (CCNFs) are used as the reinforcing material in the PLA matrix for their ability to selectively absorb microwave energy. Subsequently, there is transferal of the heat that is generated from the microwaves to the PLA matrix that triggers amorphous PLA chains to repack and convert into new crystallites. To avoid damage to the material from overheating, the composite is simultaneously cooled to mitigate excess heating from the microwaves. When comparing the microwave annealing method on a composite at a lowered temperature to the conventional oven annealing, the process time to reach the nanocomposites maximum crystallinity and ultimate tensile strength was significantly reduced.

The RF application to the AM process, in-situ was already mentioned in the previous section. However, this process also allows postprocess consolidation of printed thermosets with a much higher interfacial strength between roads and layers.[68]

Overall, this is an area that has been explored scarcely to date. While external fields are already being explored during processing, little to no work has been done to apply these techniques to post-processing, e.g., curing reactions or annealing. Specifically, annealing close to softening, as it is known from semi-crystalline polymers, liquid crystalline, block-copolymers or nanocomposites thereof offers a way of obtaining improved, higher order morphologies and densification of printed parts (e.g., physical aging).[140]

In summary, the use of morphological controls, either through external field or post processing techniques, can be used to target the nanomaterial reinforcements during the AM process, as opposed to depending on the process itself, as exhibited in the previous section. The morphological controls do not change the nature of the materials themselves, but rather the alignment or orientation of the nanoparticles within the matrix material during the AM processing. Because external fields and post processing can be utilized in conjunction with various AM methods, it allows for control over the behavior of the material itself and to be tailored to meet the specifications required for its end use. However, though these morphological controls can be applied to traditional processes as well, it is imperative to note that the coupling of these techniques with AM is mutually beneficial to the final product in a way that is unique and irreplaceable due to the nature of inherent defects in the AM process.

## 2.3 Architecture Control

The additive manufacturing process provides an additional control via print path topology on the length scale of the extrudate diameter in extrusion-based AM processes. A combination of both nanoelement alignment (shear, electric, acoustic, magnetic fields), spatial distribution (multi-material nozzles) and print path topology (print-head direction) opens a wide window of possible architectures that has previously been seen only in nature. While we understand that this section requires both specialized processes and morphologies, we believe that it merits its own section due to the nature of the higher architectures with emerging properties which are not attainable via processing and morphology alone or combined. Examples which touch on this are shape memory nanocomposites and metamaterials.

In various AM strategies, as discussed in the previous sections, controlling the nanofillers orientation and concentration can achieve complex hierarchal structures and functions in printed parts. Because architectural control involves melding elements from the previous two chapters, there are fewer research efforts dedicated to this. The work highlighted in this section exhibit the opportunity that can be found with architectural control that has yet to be realized extensively in the community.

In the seminal study of Compton and Lewis, epoxy-based inks enabled printing of different cellular composites structures of hexagonal and triangular honeycomb shapes, inspired by balsa wood. The nanocomposite inks consisted of nanoclay platelets (used as a rheology modifier), silicon carbide whiskers and short carbon fibers. The printing process induced alignment of the fillers along the print direction and the build path was used to spatially control the filler orientation within printed parts. The mechanical properties (elastic modulus and tensile strength) as a function of the relative density (density of the structure/density of the solid base material) slightly deviated from well-established scaling laws, likely due to the presence of defects and geometrical imperfections in the lattice structure, including nodal misalignment and waviness in the cell walls.[47] A way to overcome the presence of defects and prevent delamination between different printed layers in epoxy systems was demonstrated by Chen et al.[142] A formulation based on epoxy, benzoxazine and CNT showed both thermoplastic (during printing at 100 °C) and thermoset behavior (x-linking above 200 °C). A 3D flower structure was also printed with this formulation, showing a shape memory effect (i.e., opening of the flower 'pedals') with excellent recoverability (>96% of the initial angle recovered) and relatively fast recovery rate (80 s). Multiwalled carbon nanotubes were also used in the work of Hua et al. where photo-responsive shape changing nanocomposites were created on a paper substrate.[143] The polymer matrix in this case was poly(lactic acid) and the nanocomposite exhibited excellent sensitivity and photothermal effects under near-infrared irradiation. The shape-changing properties were fully reversible, and the flexible paper-based actuators were able to fully recover their original shape once the light source was switched off.

In another FDM study based on PLA filament, Shi utilized hybrid nanofillers of carbon nanotubes and graphene nanoplatelets, in combination with controlled 3D printed cellular patterns (hexagonal, square, and triangular cell geometries), to establish hierarchical structures for both electrical conductivity (82 S/m) and EMI shielding efficiency (36.8 dB at 10 GHz).[144] Various cell geometries of printed nanocomposites have been achieved, with a critical cell size identified for the maximization of the EMI shielding performance of printed structures. In another work, 2D and 3D micropatterns (i.e. grid, honeycomb, pyramid, and square) with high nanofiller loadings were successfully achieved by Wajahat et.al, who developed a hybrid nanocomposite ink based on microsized graphene flakes as the electrically conductive phase and as the supporting structure for $Fe_3O_4$ nanoparticles.[145] 3D grid structures with electrical conductivity of 580 S/m, magnetic properties of 15.8 emu/g and high EMI shielding efficiency (between 55 and 20 dB at 10 GHz, for grid pitch sizes between 1050 μm and 150 μm) were achieved.

By using magnetic field assisted 3D printing, Martin et al. recreated a series of hierarchical biologically-inspired structures, including layered nacre structures, osteon structures of cortical bone and cholesteric structures of the dactyl club of the peacock mantis shrimp.[146] The internal structure of the complex printed nanocomposite architectures was interconnected with the tensile properties and the hardness of the final materials, while the fracture mechanism could also be tuned. Even though this work stemmed from a bio-inspired phenomenon, it has opened the door for evolving the synthetically derived products have the potential to exhibit characteristics that are superior to their naturally occurring counterparts.

Yang et al. obtained an even more complicated structure with an electrically assisted 3D printing method: a human meniscus.[135] The unique circumferential and radial 3D alignment of collagen in the human meniscus is critical in preventing radial and vertical tear, respectively. This highly anisotropic and hierarchical structure was recreated by orienting MWCNT circumferentially and radially, aided by external electric fields. The so produced nanocomposites could well reproduce the meniscus structure at large scale, which, in turn, resulted in enhanced mechanical performance (e.g. tear resistance). Beyond mechanical properties, the possibility to engineer intricate nanocomposite architectures could also enable sensing ability, important in different field including bioengineering, aerospace engineering and robotics.

To utilize the synergistic architectural contributions from AM and nanofillers, Yang et al. used an electrical field to align graphene nanoplatelets in order to achieve a nacre inspired brick-mortar structure. With only 2 wt.% of highly aligned graphene fillers in printed specimens, comparable specific strength and toughness to natural nacre were obtained thanks to induced crack deflection and low density (1.06 g/cm$^3$). Anisotropic electrical properties have also been established for potential protective wearable sensors.[147]

An excellent example of how complex three-dimensional architectures obtained by additive manufacturing, can result in unique physical properties is the work of Cui et al.[148] By 3D printing piezoelectric nanocomposites slurries (surface functionalized lead zirconate titanate nanoparticle colloids cross-linked by UV curable monomers), previously inaccessible classes of piezoelectric materials were fabricated, with arbitrary piezoelectric coefficient tensors (**Figure 16**). The authors created 3D printed metamaterial blocks, starting from the inverse design of an arbitrary piezoelectric tensor, overcoming the typical coupling modes observed in piezoelectric monolithic and foams. After the polarization of the as-fabricated 3D architectures, it was demonstrated that piezoelectric behavior in any direction can be selectively reversed, suppressed, or enhanced, achieving distinct voltage response signatures with applied stress. The above design and tessellation of the piezo-active units can lead to a variety of smart-material functionalities, including vector and tactile sensing, source detection, acoustic sensing, and strain amplifications from a fraction of their parent materials.

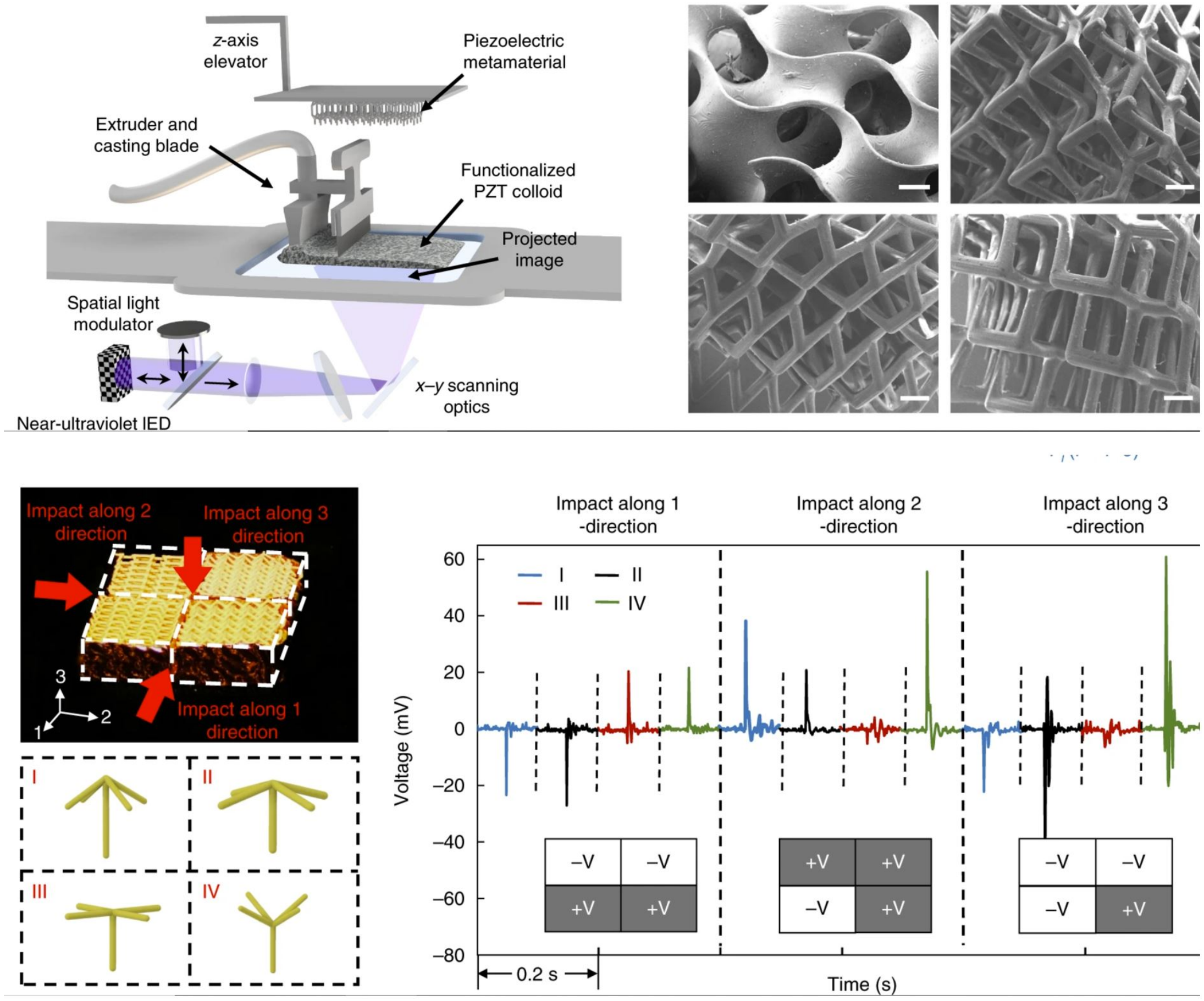

**Figure 16**. Schematic illustation of the high resolution additive manufacturing system, Scanning electron microscope images of 3D-printed piezoelectric microlattices, scale bare 300µm, as fabricated piezoelectric infrastructure comprised of stacked architectures with encoded piezoelectric constants, voltage output patterns corresponding to different impact directions indicated by red arrows. The impact force in the 1-direction is registered with permutation voltage matrix [-,-,+,+], with [+,+,-,+] for the 2 direction and [-,-,-,+] for the 3-direction, respectively. [148]

In a milestone paper, Gladman et al. printed composite hydrogel architectures inspired by botanic systems and programmed shape-changing upon immersion in water, by exploiting anisotropies in both mechanical (stiffness) and swelling behavior.[110] By combining bilayer patterns that generate simple curved surfaces (positive, negative, and varying Gaussian curvatures), the authors created a series of functional folding flower architectures, including one that resembles an orchid (**Figure 17**). When a stimuli-responsive poly(N-isopropylacrylamide) matrix was used, the same structures could also reversibly unfold upon change in temperature. The above approach was defined as 4D printing for the first time, by the same group, where the 4$^{th}$ dimension is time, as the 3D printed structure actuated and responded to a certain stimulus.

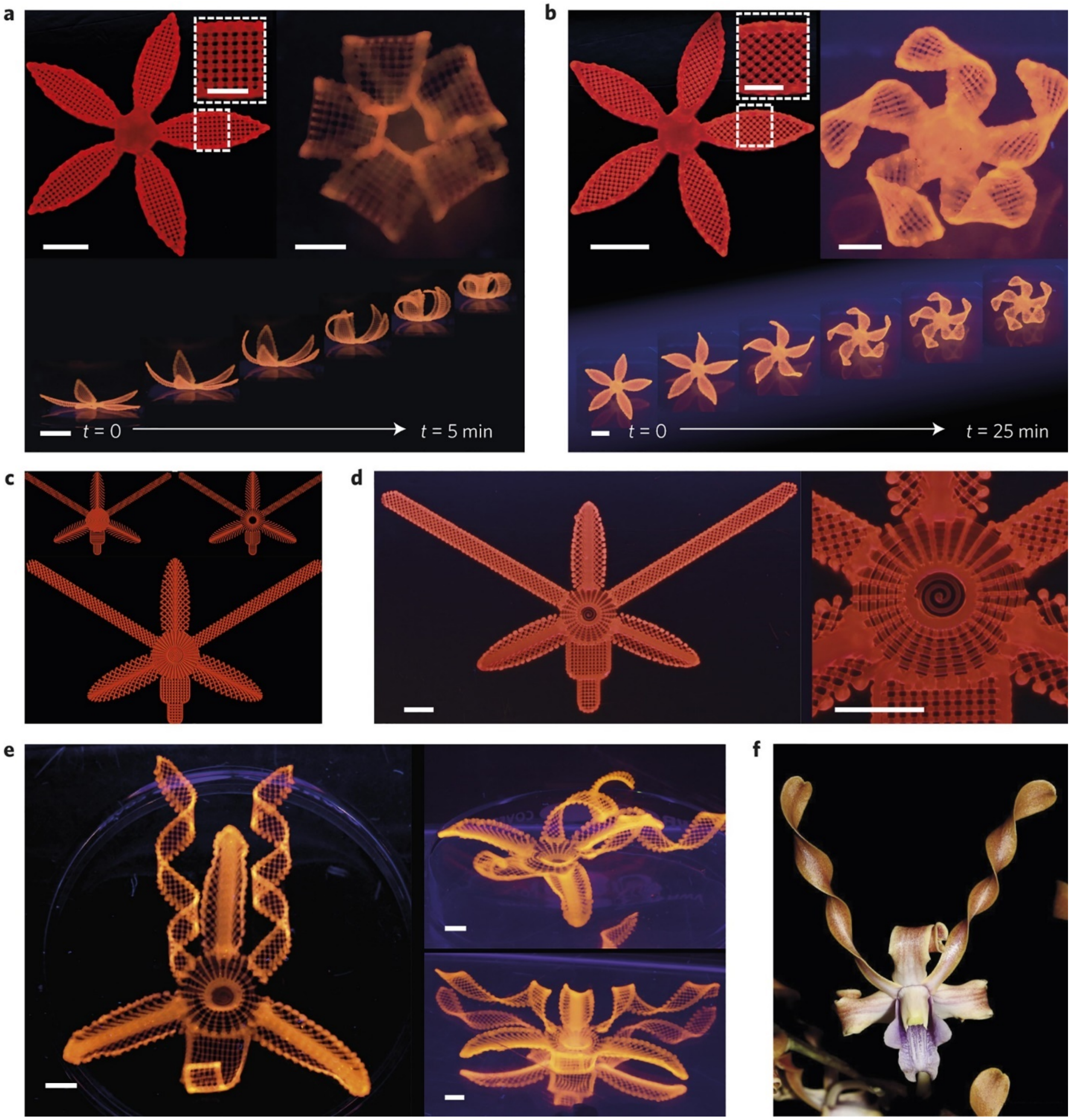

**Figure 17**: **a**,**b**, Simple flowers composed of 90°/0° (**a**) and −45°/45° (**b**) bilayers oriented with respect to the long axis of each petal, with time-lapse sequences of the flowers during the swelling process (bottom panel) (scale bars, 5 mm, inset = 2.5 mm). **c**–**f**, Print path (**c**), printed structure (**d**) and resulting swollen structure (**e**) of a flower demonstrating a range of morphologies inspired by a native orchid, the *Dendrobium helix* (courtesy of Ricardo Valentin) (**f**). Based on the print path, this orchid architecture exhibits four different configurations: bending, twisting and ruffling corolla surrounding the central funnel-like domain (scale bars, 5 mm).[110]

Several advanced technologies such as flexible electronics and soft robotics can utilize soft materials and multidomain shape-shifting properties. In a characteristic example, Kim *et al.* used an elastomer nanocomposite ink containing neodymium, iron and boron nanoparticles to achieve direct writing of soft structures.[149] The complex structures displayed multiple magnetic domains and as a response to external magnetic fields, the nanoparticles aligned themselves along the magnetic field which was smartly applied by a tunable electromagnet around the printer nozzle. The magnetically actuated multiple domains were printed successfully and the 2D and 3D structures were able to shape shift into extraordinarily complex structures. In another work, Hu and coworkers fabricated a poly(dimethylsiloxane)/Fe (PDMS/Fe) composite ink in order to 4D print structures that could evolve rapidly under an external magnetic field, as a result of the high magnetic permittivity and the low magnetic coercive force of the Fe nanoparticles.[91] The authors managed to 3D print a butterfly that could flap its wings under a magnetic field. The structure evolvements of the 3D printed structures can create functionalities for biomedical devices or smart textiles by 4D printing of this relatively simple type of ink. Similarly, Wei et al. manufactured shape memory nanocomposites consisting of poly(lactic acid) and iron oxide nanoparticles.[150] Ultraviolet crosslinking nanocomposite inks were utilized to enable the printing of 4D active shape-changing architectures. The printed objects displayed excellent shape memory behavior and the presence of iron oxide enabled the use of magnetic field for remote actuation and magnetically guidable properties. Another example of response to a magnetic field is the research of Shinoda *et al.*[151] The authors created biomimetic actuators, which were able to exhibit a metachronal wave movement, when an external rotational magnetic field was applied. The matrix in this case was an UV-curable gel and two types of ferromagnetic particles: strontium-ferrite and carbonyl iron particles. Again, the application of the external field during 3D printing led to formation of distinct magnetic domains and a well-controlled movement of a worm-type actuator and an artificial cilium.

A flexible solvent-cast 3D printing technique was employed by Wei et al. to print electrically conductive and functional parts. The integration of highly electrically conductive silver-coated carbon nanofibers (Ag@CNFs) with the shape memory behavior of the PLA polymer matrix enabled the printing of electroactive structures. As an example, a gripper triggered with a voltage as low as 1 V was successfully demonstrated.[152] Another electrically actuated flower inspired gripper was reported from Shao et al. as a result of the behavior of 4D printed polymer nanocomposites. Upon application of a voltage, thermal expansion of 4D printed nanocomposites, based on PLA and silver nanowires, took place to realize the phase change and corresponding actuation.[153]

Architectural control, as discussed, shows an ability to meld ideals from process and morphology control and, with the added variable of time, allows for structures and products to be created that show properties and behaviors unlike any of their traditionally manufactured counterparts. In some cases, described herein, it was the use of external fields that allowed for a change in the final structure and architecture of the part itself. However, it is the combination of intricate filler orientation and control, the modification of the rheological properties of the printed materials and the utilization of external fields that allows for these unique transformations to take place. Though often labelled under the umbrella of 4D printing[154, 155], the driving concepts behind the works in this field align more closely to the idea of a synergistic relationship between AM and nanocomposites than just an added “fourth dimension”. The architectural control, which is enabled by nanoparticles and AM, also opens up the design-for-manufacturing parameter space. The ability to print something that is not possible via conventional processes in fact requires a paradigm shift in the engineering design.

# 3 Combinations and Other

We tried to capture all areas that will positively affect processing, morphology, and architecture and lead to overall improvements in performance of printed parts beyond what is possible without the combination of AM and nano. We excluded new materials development for either additive manufacturing or nanocomposites or a combination of the two. These are important areas that still need to be addressed.[40] Off-the-shelf, commodity materials are good for the most rudimentary additive manufacturing processes, which include complex shaped tools or prototyping for fit, form and function. However, true advantages will only be enabled when new matrix materials are developed that adapt to the processing conditions, such as FDM or SLS. Several groups use nanoparticles that both improve or enable processing, while also contributing functionality in the printed part. A few examples are the work by Erb et al. and Gelinsky et. al who used Laponite as a processing aid while the Laponite also increased shape fidelity and was beneficial for controlled release of biologically active agents.[132, 156] An interesting approach to new morphologies with nanoscale porosity is summarized in a review by Fujita et. al in which 3D printing is coupled to a de-alloying process to a second nanophase which is separated from the printed part to obtain hierarchical structures that are otherwise not possible.[157] A few articles discuss process modeling efforts to understand the additive manufacturing process with nanoparticles. Gu et. al describe a multiscale computational numerical modeling framework that may become an important tool in predicting thermodynamic and kinetic mechanisms during SLM AM/3DP processes with nanoparticle contributions.[49] Kim et. al study the thermodynamic behavior of powders during laser heating and apply molecular dynamics simulations to study Ag nanoclusters (2-7nm) during heating and cooling process and find a mesoscale regime where properties of nanoclusters can be described with macroscopic concepts.[71]

## 4 Outlook & Opportunities

Additive manufacturing (AM) has been developed for several decades, but the technology has only received wide attention in the last 10 years with the surge in publications, patents and reports with novel ideas, concepts and engineering solutions that allowed almost anybody to do AM in their own laboratory. Once the first inexpensive 3D printers were commercially available, researchers started printing everything from gels to ceramics to filled polymers and a large part of the community quickly lost sight of the true capabilities of AM, which includes rapid prototyping, part consolidation and complexity enabled capabilities. The latter two require more than just adding AM and nano – it necessitates cooperative effects between the two to accomplish the goals.

### 4.1 Challenges

Each of the technologies in their own field have challenges still to overcome. The combination of nanocomposites with additive manufacturing in the implementation phase causes these challenges to magnify. One of these limitations is the perception that a simple combination of a polymer with a particle will result in a novel nanocomposite. It took 20 years for the polymer nanocomposite community to realize that it requires a combination of the right chemistry (compatibility) and mechanical assistance (external fields, shear) to obtain materials with superior mechanical and electrical properties over simple physically mixed composites.[139] Today, it is well known that the selection of the polymer and nanofillers highly depends on the desired properties for a targeted application, operation window and processing/fabrication technique. Additive manufacturing with its many different engineering solutions has process-structure-property requirements which, at the extremely basic level, add more constraints to nanocomposite formation and processing. Hence, learning how the process affects the structure/texture is crucial in obtaining tailored structures and thus achieve the desired mechanical and functional properties. Implementation of novel in situ metrology and in-operando characterization of materials during the AM process with the ability for atomic/nano-scale resolution are promising pathways in understanding and tailoring the formation of structure during the process. In this view, multiscale modelling will be an essential tool to complement our understanding in linking process-structure-property. Many conceptual ideas have not been realized yet and the opportunity field is still wide open in terms of complexity enabled capabilities that AM makes available. This article presents a few successful laboratory examples in which these two technologies have been combined into products that take advantage of both the presence of nanoparticles and the AM processing, leading to synergistic enhancements in processing and properties of the resulting coupons, parts, and components. This includes development of new resin materials with chemistries that adapt to the different AM engineering solutions and processing routes.

Additionally, these AM processes are continuing to be developed and perfected at a different rate than their material counterparts; moreover, they are not being developed in parallel or synergistically. Many of these advancements are made on a laboratory scale and were not intended to scale into a mass production setting. The separate paths of improving AM versus improving nanocomposites are likely a result of a knowledge gap between the designers of an AM process and the developers of a nanocomposite material, where they are not being optimized to complement each other, but instead stand alone.

Other examples which illustrate why AM + Nano has not been realized and implemented yet in industrial processes include commodity polymers such as polypropylene. These polymers were targeted by the nanocomposite community because of their wide-spread utilization in many industrial and day-to-day applications, ranging from electronics (BOPP), food packaging and automotive industry.[158] The additive manufacturing of polypropylene is not trivial due to high melt viscosities and the challenges with the addition of functional nanoparticles (non-polar matrix). Only recently jet fusion was developed as a robust engineering solution to address this challenge for polypropylene.[159]

Furthermore, health implications prevented early adoption for polymer nanocomposites. Adding nanoparticulates to a novel process such as AM in the laboratory is not an issue. However, this changes drastically when applied into an industrial process that leads to distribution of articles and parts into widespread use. There is also the perception and in some cases evidence that nanoparticles can lead to long-term illnesses such as cancer. This on the other hand leads to regulatory aspects which prevent immediate adoption in large scale manufacturing.

## 4.2 Realization

As the nanocomposite community jumped onto the AM publishing bandwagon and a race started to publish on anything that had nanoparticles in them it was evident that most of these publications only demonstrated extrusion of nanocomposites and used the additive manufacturing moniker for the sake of publishing. This was also the case for the emergence of nanocomposites at the end of the 1980's when researchers added any known nanoparticle to a host of polymers until realization set in that it is not just a simple process of adding two parts. A primary concern is that this will lead to frustration and disillusionment and will, in turn, negatively affect the progress of AM with poor performance and defect ridden morphologies of the resulting printed articles with poor acceptance and buy-in from industry and it can take years to regain the trust of industry. This was evident during the writing of this article with fewer than 50% of several thousand articles fulfilling our criteria of synergistic use of AM + Nano. A solid understanding of the process and impactful integration of nanoparticles during the AM or 3D printing process is necessary.

### *4.2.1 Recent Examples*

While the incorporation of carbon black, chopped carbon fibers and continuous carbon fibers has been widely adopted by industry for AM, nanomaterials for AM is still an emerging field with small businesses taking a leap in this area. This is notably evident in the area of thermosetting resins for which novel thermal curing mechanisms are proposed that include nanoparticles for converting the heat of an external source (UV, laser, RF, plasma, micro wave, acoustic, IR, magnetic) into a curing exotherm during the AM process or for the manufacture of functional composites (conductivity).[160-165]

With the development of conductive inks in areas such as RFID antennas and flexible circuits, UV curable resins are used.[160, 162, 165] Laser sources are used to consolidate materials into printed antennas for next generation wireless devices and thermal protection materials and internal insulation more recently.[161, 162] A unique route was realized in the case of magnetic field alignment during processing, which allows the manufacture of 3D printed RF lenses, antennas and wave guide designs which cannot be produced using conventional manufacturing routes.[164] The application of RF leads to plasma enables consolidation and of jigs and fixtures for light weighting and tooling.[165]

#### *4.2.2 Potential*

The sky is the limit.

Impactful progress has been made that leads to the first emerging small start-up and larger companies and OEMs embracing AM with the inclusion of nanoparticles. For example, solicitation for proposals from small businesses to military applications by the Army and the Air Force are a testament for the change in the hype curve. Industry and the Department of Defense realize the potential of these materials to create a more agile manufacturing foundation not just limited to niche applications.

As the automotive and aerospace sectors identify the benefits from these initial technology transitions, AM of nanocomposite will soon form a stronger presence (~3-5 years). This is particularly true with the rise in urban air mobility technology developments, which require complex, light weight structures and energy sources.[166] Similarly, the Air Force is pivoting towards limited life technologies and is investing in attainable and smaller, low-cost aircraft technologies that require condensed certification protocols.[167] These disruptive technologies will require a riskier approach than in the past to go from concept to manufacturing. AM of nanocomposites will naturally be a part of that change in how people will conduct urban travel and how the DoD will support the warfighter.[168]

Long-term technology transitions that are enabled by AM of nanocomposites are areas of structural health monitoring, in situ sensing during the AM process and large area shielding applications. Even longer term, the AM of external field induced self-assembly of nanoparticles into fibrillar, and continuous structures could be used for reinforcement and supplement the current surge in continuous carbon fiber AM.

The next phase of AM of nanocomposites needs to embrace the full spectrum of digital manufacturing which must include topology optimization to drive local directionality and concentration of nanoparticles. The full potential of hierarchical architectures can then be realized for mechanical and electromagnetic requirements of complex parts.

Many of the extreme requirements that will enable human travel to Mars will take advantage of these new emerging technologies because conventional materials do not exhibit the energy densities of topology optimized composites. NASA is already funding companies to 3D print in space.[169] Polymers need to be protected from the extreme environment and addition of nanoparticles is one solution to this challenge.

The universe is the limit.

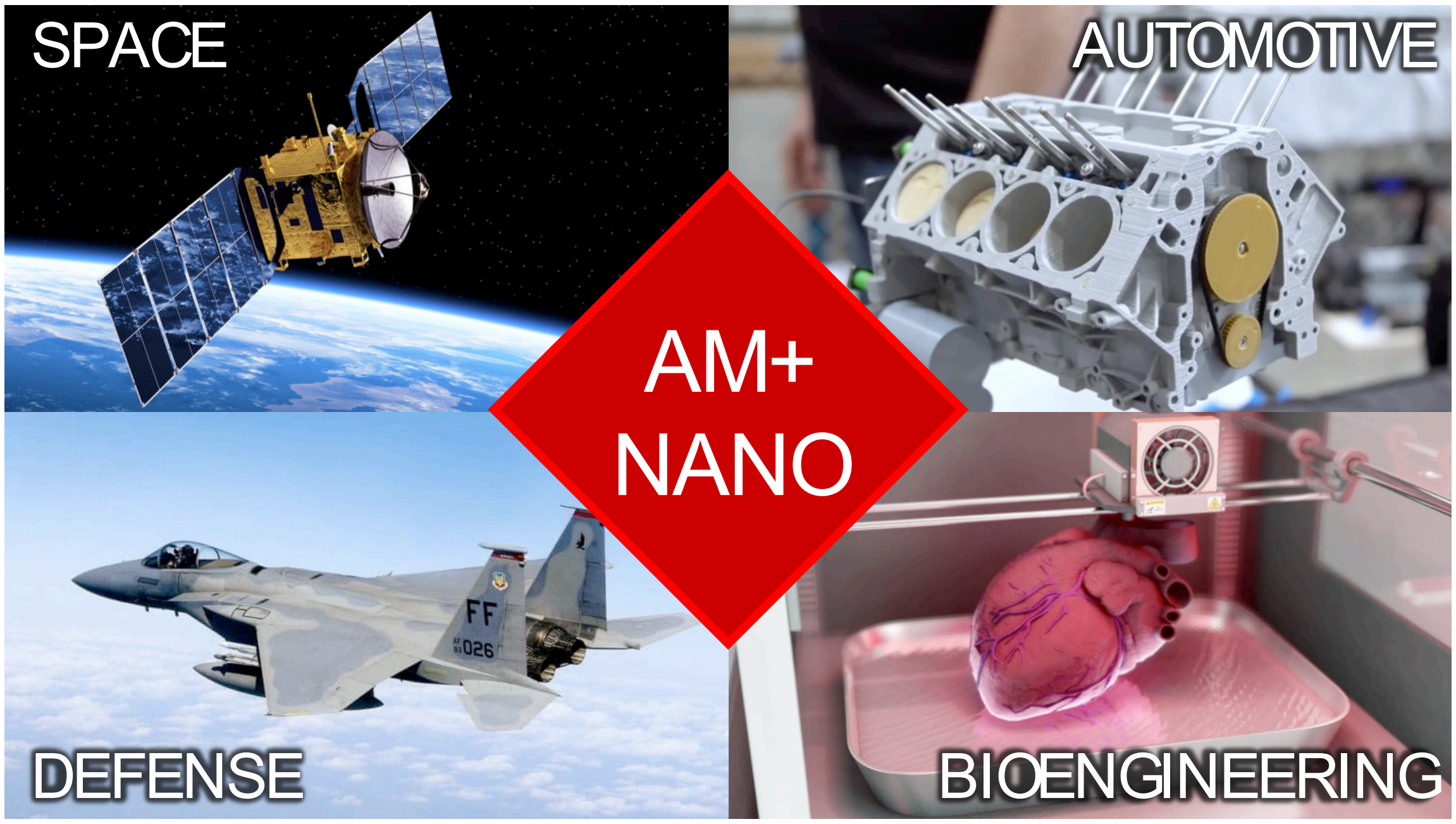

**Figure 18**. Applications leveraging additive manufacturing with nanoparticles, including a) satellites, commercial drones and flying taxis; b) defense applications such as unmanned aerial vehicles and sustainment of existing fleets; c) automotive and civil engineering sectors, such as structural cement walls for buildings and homes; and d) individualized bone grafts, organs and prosthetics,.[170-173]

## 5. Acknowledgements

DGP, EB and HZ acknowledge the support from "Graphene Core 3" GA: 881603 which is implemented under the EU-Horizon 2020 Research & Innovation Actions (RIA) and is financially supported by EC-financed parts of the Graphene Flagship. MC acknowledges support from the Air Force Research Laboratory Scholarship program. HK acknowledges support from the Air Force Office of Scientific Research (M.-J. Pan) under the Low Density Portfolio # 17RXCOR436. The authors would like to thank Kyle Johnson and Ruel McKenzie for their help with literature.

## References

[1] I. Gibson, D. W. Rosen, B. Stucker, Springer US, US 2015.

[2] A. C. de Leon, Q. Chen, N. B. Palaganas, J. O. Palaganas, J. Manapat, R. C. Advincula, *Reactive and Functional Polymers* **2016**, *103*, 141.

[3] (Ed: I. AREVO), AREVO, Inc., 2017.

[4] (Ed: C. Composites), Continuous Composites, 2018.

[5] J. Vurpillat, (Ed: Stratasys), 2017.

[6] C. M. Stokes-Griffin, P. Compston, *Composites Part A: Applied Science and Manufacturing* **2015**, *75*, 104.

[7] Vol. 2018 (Ed: I. Objects), Impossible Objects, 2017.

[8] S. R. Shin, R. Farzad, A. Tamayol, V. Manoharan, P. Mostafalu, Y. S. Zhang, M. Akbari, S. M. Jung, D. Kim, M. Comotto, N. Annabi, F. E. Al-Hazmi, M. R. Dokmeci, A. Khademhosseini, *Advanced Materials* **2016**, *28*, 3280.

[9] R. Rahimi, M. Ochoa, B. Ziaie, *ACS Applied Materials and Interfaces* **2016**, *8*, 16907.

[10] A. E. Jakus, A. L. Rutz, R. N. Shah, *Biomed Mater* **2016**, *11*, 014102.

[11] C. K. Chua, W. Y. Yeong, J. An, *Molecules* **2016**, *21*.

[12] C. M. O'Brien, B. Holmes, S. Faucett, L. G. Zhang, *Tissue Eng., Part B* **2015**, *21*, 103.

[13] M. McAlpine, "3D printed bionic nanomaterials", 2016.

[14] C. Jae-Won, M. Vatani, E. D. Engeberg, "Direct-write of multi-layer tactile sensors", presented at *2013 13th International Conference on Control, Automaton and Systems (ICCAS 2013), 20-23 Oct. 2013*, Piscataway, NJ, USA, 2013.

[15] C. C. Spackman, C. R. Frank, K. C. Picha, J. Samuel, *Journal of Manufacturing Processes* **2016**, *23*, 296.

[16] J. M. Gardner, G. Sauti, J.-W. Kim, R. J. Cano, R. A. Wincheski, C. J. Stelter, B. W. Grimsley, D. C. Working, E. J. Siochi, *Additive Manufacturing* **2016**, *12, Part A*, 38.

[17] P. Colombo, E. Bernardo, G. Parcianello, *J. Eur. Ceram. Soc.* **2013**, *33*, 453.

[18] M. A. Ganter, D. W. Storti, G. Marchelli, A. P. K. Gramling, "3D printing of a glass-ceramic composite", presented at *SAMPE Tech Seattle 2014 Conference, June 2, 2014 - June 5, 2014*, 800 Convention Place, Seattle, WA 98101-2350, United states, 2014.

[19] M. Lorusso, A. Aversa, D. Manfredi, F. Calignano, E. P. Ambrosio, D. Ugues, M. Pavese, *J. Mater. Eng. Perform.* **2016**, *25*, 3152.

[20] S. Cao, D. Gu, *J. Mater. Res.* **2015**, *30*, 3616.

[21] Q. Jia, D. Gu, *J. Mater. Res.* **2014**, *29*, 1960.

[22] P. Wang, B. Zhang, C. C. Tan, S. Raghavan, Y.-F. Lim, C.-N. Sun, J. Wei, D. Chi, *Materials & Design* **2016**, *112*, 290.

[23] O. Ivanova, C. Williams, T. Campbell, *Rapid Prototyping Journal* **2013**, *19*, 353.

[24] R. D. Farahani, M. Dubé, D. Therriault, *Advanced Materials* **2016**, *28*, 5794.

[25] K. Zhong Xun, J. Ee Mei Teoh, L. Yong, C. Chee Kai, Y. Shoufeng, A. Jia, L. Kah Fai, Y. Wai Yee, *Virtual and Physical Prototyping* **2015**, *10*, 103.

[26] V. Francis, P. K. Jain, *International Journal of Rapid Manufacturing* **2015**, *5*, 215.

[27] J.-Y. Choi, S. Das, N. D. Theodore, I. Kim, C. Honsberg, H. W. Choi, T. L. Alford, *ECS Journal of Solid State Science and Technology* **2015**, *4*, P3001.

[28] D. S. Engstrom, B. Porter, M. Pacios, H. Bhaskaran, *J. Mater. Res.* **2014**, *29*, 1792.

[29] R. Vaidyanathan, in *Additive Manufacturing*, CRC Press, 2015, 19.

[30] E. P. Koumoulos, E. Gkartzou, C. A. Charitidis, *Manufacturing Rev.* **2017**, *4*, 12.

[31] J. Z. Manapat, Q. Chen, P. Ye, R. C. Advincula, *Macromolecular Materials and Engineering* **2017**, *302*, 1600553.
[32] J. Li, S. G. Ballmer, E. P. Gillis, S. Fujii, M. J. Schmidt, A. M. E. Palazzolo, J. W. Lehmann, G. F. Morehouse, M. D. Burke, *Science* **2015**, *347*, 1221.
[33] NIST, NIST, Gaithersburg, MD 20899 2016, 57.
[34] J. Kapustin, D. Braley, T. Whitney, T. Santelle, C. Patterson, B. Czapor, in *DTIC*, DTIC, 2015.
[35] T. Albers, 2017.
[36] P. Song, Z. Fu, L. Liu, C.-W. Fu, *Computer Aided Geometric Design* **2015**, *35–36*, 137.
[37] T. H. Ware, J. S. Biggins, A. F. Shick, M. Warner, T. J. White, *Nature Communications* **2016**, *7*, 10781.
[38] E. M. National Academies of Sciences, P. G. Affairs, B. I. S. Organizations, U. S. N. C. T. A. Mechanics, M. Schwalbe, *Predictive Theoretical and Computational Approaches for Additive Manufacturing: Proceedings of a Workshop*, National Academies Press, 2016.
[39] ASTM, ASTM, West Conshohocken, PA 2012.
[40] S. C. Ligon, R. Liska, J. Stampfl, M. Gurr, R. Mulhaupt, *Chemical reviews* **2017**, *117*, 10212.
[41] V. Mittal, in *Manufacturing of Nanocomposites with Engineering Plastics*, Woodhead Publishing, 2015, 15.
[42] B. H. King, D. Dimos, P. Yang, S. L. Morissette, *Journal of Electroceramics* **1999**, *3*, 173.
[43] M. P. Avery, S. Klein, R. Richardson, P. Bartlett, G. Adams, F. Dickin, S. Simske, "The rheology of dense colloidal pastes used in 3D-printing", presented at *30th International Conference on Digital Printing Technologies and Digital Fabrication 2014, NIP 2014, September 7, 2014 - September 11, 2014*, Philadelphia, PA, United states, 2014.
[44] R. McKenzie, H. Koerner, "Hierarchically Reinforced Epoxy Based Functional Nanocomposites", presented at *Materials Science & Technology 2016*, Salt Lake City, Utah, 2016.
[45] G. Postiglione, G. Natale, G. Griffini, M. Levi, S. Turri, *Polym. Compos.* **2015**, Ahead of Print.
[46] J. A. Lewis, B. G. Compton, J. R. Raney, T. J. Ober, *WO2015120429A1*, 2015.
[47] B. G. Compton, J. A. Lewis, *Adv. Mater. (Weinheim, Ger.)* **2014**, *26*, 5930.
[48] S. Hong, D. Sycks, H. F. Chan, S. Lin, G. P. Lopez, F. Guilak, K. W. Leong, X. Zhao, *Adv. Mater. (Weinheim, Ger.)* **2015**, *27*, 4035.
[49] D. Gu, C. Ma, M. Xia, D. Dai, Q. Shi, *Engineering* **2017**, *3*, 675.
[50] L. M. Guiney, N. D. Mansukhani, A. E. Jakus, S. G. Wallace, R. N. Shah, M. C. Hersam, *Nano letters* **2018**, *18*, 3488.
[51] Y. Jin, A. Compaan, W. Chai, Y. Huang, *ACS applied materials & interfaces* **2017**, *9*, 20057.
[52] F.-Y. Hsieh, H.-H. Lin, S.-h. Hsu, *Biomaterials* **2015**, *71*, 48.
[53] S. Unterman, L. F. Charles, S. E. Strecker, D. Kramarenko, D. Pivovarchik, E. R. Edelman, N. Artzi, *ACS Nano* **2017**.
[54] B. Wang, A. J. Benitez, F. Lossada, R. Merindol, A. Walther, *Angewandte Chemie - International Edition* **2016**, *55*, 5966.
[55] A. Tang, Y. Liu, Q. Wang, R. Chen, W. Liu, Z. Fang, L. Wang, *Carbohydr. Polym.* **2016**, *148*, 29.
[56] K. Haakansson, C. de la Pena, V. Kuzmenko, P. Enoksson, P. Gatenholm, "3D printing of a conductive ink based on cellulose nanofibrils and carbon nanotubes", 2016.
[57] A. Rees, L. C. Powell, G. C. Carrasco, D. T. Gethin, K. Syverud, K. E. Hill, D. W. Thomas, *BioMed Res. Int.* **2015**, 1.
[58] K. Markstedt, A. Mantas, I. Tournier, H. Martinez Avila, D. Haegg, P. Gatenholm, *Biomacromolecules* **2015**, *16*, 1489.
[59] C. B. Highley, C. B. Rodell, J. A. Burdick, *Advanced Materials* **2015**, *27*, 5075.
[60] K. Haakansson, I. Henriksson, P. Gatenholm, "Solidification of 3D printed cellulose nanofibrils", 2016.

[61] J. Linkhorst, K. Percin, S. Kriescher, M. Wessling, *ChemElectroChem* **2017**, *4*, 2760.
[62] D. Kokkinis, M. Schaffner, A. R. Studart, *Nature Communications* **2015**, *6*, 8643 (10 pp.).
[63] Z. Weng, Y. Zhou, W. Lin, T. Senthil, L. Wu, *Composites, Part A* **2016**, *88*, 234.
[64] L. Yihua, A. E. H. Charlotte, *Biomedical Materials* **2016**, *11*, 014103.
[65] K. Yadav, M. Jassal, A. K. Agrawal, *Journal of the American Ceramic Society* **2014**, *97*, 4031.
[66] J. Schmidt, M. Sachs, C. Blümel, B. Winzer, F. Toni, K.-E. Wirth, W. Peukert, *Procedia Engineering* **2015**, *102*, 550.
[67] Y. Zhang, W. R. Schmidt, *WO2015112366A1*, 2015.
[68] C. B. Sweeney, M. J. Green, M. Saed, *WO2015130401A2*, 2015.
[69] C. B. Sweeney, B. A. Lackey, M. J. Pospisil, T. C. Achee, V. K. Hicks, A. G. Moran, B. R. Teipel, M. A. Saed, M. J. Green, *Science Advances* **2017**, *3*, e1700262.
[70] D. A. Champion, J. E. Abbott, Jr., *WO2016118151A1*, 2016.
[71] T. Q. Vo, B. H. Kim, *International Journal of Precision Engineering and Manufacturing-Green Technology* **2017**, *4*, 301.
[72] C. E. Duty, V. Kunc, L. J. Love, W. H. Peter, O. Rios, *US20150183138A1*, 2015.
[73] R. Erb, J. J. Martin, *WO2015188175A1*, 2015.
[74] M. Lau, R. G. Niemann, M. Bartsch, W. O'Neill, S. Barcikowski, *Appl. Phys. A: Mater. Sci. Process.* **2014**, *114*, 1023.
[75] K. Schuetz, J. Hoerber, J. Franke, *AIP Conf. Proc.* **2014**, *1593*, 732.
[76] A. A. Pawar, S. Halivni, N. Waiskopf, Y. Ben-Shahar, M. Soreni-Harari, S. Bergbreiter, U. Banin, S. Magdassi, *Nano Lett* **2017**, *17*, 4497.
[77] J. Mendez-Ramos, J. C. Ruiz-Morales, P. Acosta-Mora, N. M. Khaidukov, *Journal of Materials Chemistry C* **2016**, *4*, 801.
[78] D. Lin, M. Motlag, M. Saei, S. Jin, R. M. Rahimi, D. Bahr, G. J. Cheng, *Acta Materialia* **2018**, *150*, 360.
[79] P. Rosso, L. Ye, *Macromolecular Rapid Communications* **2007**, *28*, 121.
[80] C. Chen, R. S. Justice, D. W. Schaefer, J. W. Baur, *Polymer* **2008**, *49*, 3805.
[81] G. Kalay, M. J. Bevis, in *Polypropylene: An A-Z reference*, (Ed: J. Karger-Kocsis), Springer Netherlands, Dordrecht 1999, 38.
[82] J.-D. Cho, H.-T. Ju, J.-W. Hong, *Journal of Polymer Science Part A: Polymer Chemistry* **2005**, *43*, 658.
[83] C. Check, R. Chartoff, S. Chang, *European Polymer Journal* **2015**, *70*, 166.
[84] A. A. Pawar, G. Saada, I. Cooperstein, L. Larush, S. Magdassi, J. A. Jackman, S. R. Tabaei, N.-J. Cho, *Sci Adv* **2016**, *2*, e1501381.
[85] Y. Han, F. Wang, C. Y. Lim, H. Chi, D. Chen, F. Wang, X. Jiao, *ACS applied materials & interfaces* **2017**, *9*, 32418.
[86] J. H. Martin, B. D. Yahata, J. M. Hundley, J. A. Mayer, T. A. Schaedler, T. M. Pollock, *Nature* **2017**, *549*, 365.
[87] A. Chiappone, E. Fantino, I. Roppolo, M. Lorusso, D. Manfredi, P. Fino, C. F. Pirri, F. Calignano, *ACS Appl. Mater. Interfaces* **2016**, *8*, 5627.
[88] I. S. Kim, S. R. Kim, S. H. Son, J. W. Han, S. Y. An, *US20120138215A1*, 2012.
[89] E. Fantino, A. Chiappone, I. Roppolo, D. Manfredi, R. Bongiovanni, C. F. Pirri, F. Calignano, *Adv. Mater. (Weinheim, Ger.)* **2016**, *28*, 3712.
[90] H. Ma, J. Luo, Z. Sun, L. Xia, M. Shi, M. Liu, J. Chang, C. Wu, *Biomaterials* **2016**, *111*, 138.
[91] Q. Hu, X.-Z. Sun, C. D. J. Parmenter, M. W. Fay, E. F. Smith, G. A. Rance, Y. He, F. Zhang, Y. Liu, D. Irvine, C. Tuck, R. Hague, R. Wildman, *Scientific Reports* **2017**, *7*, 17150.
[92] G. Taormina, C. Sciancalepore, F. Bondioli, M. Messori, *Polymers* **2018**, *10*, 212.
[93] K. Geng, S. Li, Y. Yang, R. Misra, *Carbon* **2020**.

[94] C. Cai, C. Radoslaw, J. Zhang, Q. Yan, S. Wen, B. Song, Y. Shi, *Powder technology* **2019**, *342*, 73.
[95] R. Reese, H. Bheda, *US20160297935A1*, 2016.
[96] Y. Han, C. C. J. Yeo, D. Chen, F. Wang, Y. Chong, X. Li, X. Jiao, F. Wang, *New Journal of Chemistry* **2017**, *41*, 8407.
[97] J. H. Kim, W. S. Chang, D. Kim, J. R. Yang, J. T. Han, G.-W. Lee, J. T. Kim, S. K. Seol, *Adv. Mater. (Weinheim, Ger.)* **2015**, *27*, 157.
[98] C. Kullmann, N. C. Schirmer, M.-T. Lee, S. H. Ko, N. Hotz, C. P. Grigoropoulos, D. Poulikakos, *J. Micromech. Microeng.* **2012**, *22*, 055022/1.
[99] Y. Liu, Q. Hu, F. Zhang, C. Tuck, D. Irvine, R. Hague, Y. He, M. Simonelli, G. A. Rance, E. F. Smith, R. D. Wildman, *Polymers* **2016**, *8*.
[100] M. Layani, I. Cooperstein, S. Magdassi, *Journal of Materials Chemistry C* **2013**, *1*, 3244.
[101] J. Li, F. Rossignol, J. Macdonald, *Lab on a Chip* **2015**, *15*, 2538.
[102] M. Lee, H.-Y. Kim, *Langmuir* **2014**, *30*, 1210.
[103] B. Yongxiao, H. Szushen, N. A. Kotov, *Nanoscale* **2012**, *4*, 4393.
[104] M. Breitwieser, C. Klose, M. Klingele, A. Hartmann, J. Erben, H. Cho, J. Kerres, R. Zengerle, S. Thiele, *Journal of Power Sources* **2017**, *337*, 137.
[105] H. Khare, N. Gosvami, I. Lahouij, Z. Milne, J. McClimon, R. W. Carpick, *Nano letters* **2018**, *18*, 6756.
[106] A. Golbang, E. Harkin-Jones, M. Wegrzyn, G. Campbell, E. Archer, A. McIlhagger, *Additive Manufacturing* **2020**, *31*, 100920.
[107] A. K. Rajasekharan, R. Bordes, C. Sandstrom, M. Ekh, M. Andersson, *Small* **2017**, *13*.
[108] W. Zhao, S. Hu, Z. Shi, T. Santaniello, C. Lenardi, J. Huang, *Composites Part A: Applied Science and Manufacturing* **2020**, *129*, 105707.
[109] Y. Jin, Y. Shen, J. Yin, J. Qian, Y. Huang, *ACS applied materials & interfaces* **2018**, *10*, 10461.
[110] A. Sydney Gladman, E. A. Matsumoto, R. G. Nuzzo, L. Mahadevan, J. A. Lewis, *Nat Mater* **2016**, *15*, 413.
[111] J. T. Muth, D. M. Vogt, R. L. Truby, Y. Menguc, D. B. Kolesky, R. J. Wood, J. A. Lewis, *Advanced Materials* **2014**, *26*, 6307.
[112] T. Kim, R. Trangkanukulkij, W. S. Kim, *Scientific reports* **2018**, *8*, 1.
[113] E. A. Papon, A. Haque, *Journal of Reinforced Plastics and Composites* **2018**, *37*, 381.
[114] D. Mondal, T. L. Willett, *Journal of the Mechanical Behavior of Biomedical Materials* **2020**, *104*, 103653.
[115] M. H. Malakooti, F. Julé, H. A. Sodano, *ACS applied materials & interfaces* **2018**, *10*, 38359.
[116] Z. Liang, Y. Pei, C. Chen, B. Jiang, Y. Yao, H. Xie, M. Jiao, G. Chen, T. Li, B. Yang, *ACS nano* **2019**, *13*, 12653.
[117] R. Haney, P. Tran, E. B. Trigg, H. Koerner, T. Dickens, S. Ramakrishnan, *Additive Manufacturing* **2021**, *37*, 101618.
[118] G. Siqueira, D. Kokkinis, R. Libanori, M. K. Hausmann, A. S. Gladman, A. Neels, P. Tingaut, T. Zimmermann, J. A. Lewis, A. R. Studart, *Advanced Functional Materials* **2017**, 1604619.
[119] B. AlMangour, D. Grzesiak, J.-M. Yang, *Journal of Alloys and Compounds* **2017**, *728*, 424.
[120] M. Chen, X. Li, G. Ji, Y. Wu, Z. Chen, W. Baekelant, K. Vanmeensel, H. Wang, J.-P. Kruth, *Applied Sciences* **2017**, *7*, 250.
[121] K. J. Johnson, L. Wiegart, A. C. Abbott, E. B. Johnson, J. W. Baur, H. Koerner, *Langmuir* **2019**, *35*, 8758.
[122] M. Torres Arango, Y. Zhang, C. Zhao, R. Li, G. Doerk, D. Nykypanchuk, Y. c. K. Chen-Wiegart, A. Fluerasu, L. Wiegart, *Materials Today Physics* **2020**, *14*, 100220.
[123] Y. Shmueli, Y.-C. Lin, X. Zuo, Y. Guo, S. Lee, G. Freychet, M. Zhernenkov, T. Kim, R. Tannenbaum, G. Marom, *Composites Science and Technology* **2020**, 108227.

[124] B. P. Croom, A. Abbott, J. W. Kemp, L. Rueschhoff, L. Smieska, A. Woll, S. Stoupin, H. Koerner, *Additive Manufacturing* **2021**, *37*, 101701.
[125] E. B. Trigg, N. S. Hmeidat, L. M. Smieska, A. R. Woll, B. G. Compton, H. Koerner, *Additive Manufacturing* **2021**, *37*, 101729.
[126] J. S. Oakdale, J. Ye, W. L. Smith, J. Biener, *Opt. Express* **2016**, *24*, 27077.
[127] L. Yanfeng, M. Vatani, C. Jae-Won, *Journal of Mechanical Science and Technology* **2013**, *27*, 2929.
[128] J. Che, K. Park, C. A. Grabowski, A. Jawaid, J. Kelley, H. Koerner, R. A. Vaia, *Macromolecules (Washington, DC, U. S.)* **2016**, *49*, 1834.
[129] J.-F. Pan, S. Li, C.-A. Guo, D.-L. Xu, F. Zhang, Z.-Q. Yan, X.-M. Mo, *Sci. Technol. Adv. Mater.* **2015**, *16*, 045001/1.
[130] Q. Zhang, F. Zhang, S. P. Medarametla, H. Li, C. Zhou, D. Lin, *Small* **2016**, *12*, 1702.
[131] D. Van der Beek, A. Petukhov, P. Davidson, J. Ferré, J. Jamet, H. Wensink, G. Vroege, W. Bras, H. Lekkerkerker, *Physical Review E* **2006**, *73*, 041402.
[132] R. M. Erb, R. Libanori, N. Rothfuchs, A. R. Studart, *Science (Washington, DC, U. S.)* **2012**, *335*, 199.
[133] R. Tognato, A. R. Armiento, V. Bonfrate, R. Levato, J. Malda, M. Alini, D. Eglin, G. Giancane, T. Serra, *Advanced Functional Materials* **2019**, *29*, 1804647.
[134] L. Lu, P. Guo, Y. Pan, *Journal of Manufacturing Science and Engineering* **2017**, *139*, 071008.
[135] Y. Yang, Z. Chen, X. Song, Z. Zhang, J. Zhang, K. K. Shung, Q. Zhou, Y. Chen, *Advanced Materials* **2017**, *29*, 1605750.
[136] L. Friedrich, R. Collino, T. Ray, M. Begley, *Sensors and Actuators A: Physical* **2017**, *268*, 213.
[137] S. Bodkhe, P. S. Rajesh, F. P. Gosselin, D. Therriault, *ACS Applied Energy Materials* **2018**, *1*, 2474.
[138] S. Shariatnia, A. V. Kumar, O. Kaynan, A. Asadi, *ACS Applied Nano Materials* **2020**.
[139] B. Elder, R. Neupane, E. Tokita, U. Ghosh, S. Hales, Y. L. Kong, *Advanced Materials* **2020**, *32*, 1907142.
[140] H. Koerner, E. Opsitnick, C. A. Grabowski, L. F. Drummy, M.-S. Hsiao, J. Che, M. Pike, V. Person, M. R. Bockstaller, J. S. Meth, R. A. Vaia, *Journal of Polymer Science Part B: Polymer Physics* **2016**, *54*, 319.
[141] J. Dong, X. Huang, P. Muley, T. Wu, M. Barekati-Goudarzi, Z. Tang, M. Li, S. Lee, D. Boldor, Q. Wu, *Composites Part B: Engineering* **2020**, *184*, 107640.
[142] Q. Chen, L. Han, J. Ren, L. Rong, P. Cao, R. C. Advincula, *ACS applied materials & interfaces* **2020**, *12*, 50052.
[143] D. Hua, X. Zhang, Z. Ji, C. Yan, B. Yu, Y. Li, X. Wang, F. Zhou, *Journal of Materials Chemistry C* **2018**, *6*, 2123.
[144] S. Shi, Z. Peng, J. Jing, L. Yang, Y. Chen, *ACS Sustainable Chemistry & Engineering* **2020**, *8*, 7962.
[145] M. Wajahat, J. H. Kim, J. Ahn, S. Lee, J. Bae, J. Pyo, S. K. Seol, *Carbon* **2020**, *167*, 278.
[146] J. J. Martin, B. E. Fiore, R. M. Erb, *Nature Communications* **2015**, *6*, 8641 (7 pp.).
[147] Y. Yang, X. Li, M. Chu, H. Sun, J. Jin, K. Yu, Q. Wang, Q. Zhou, Y. Chen, *Science advances* **2019**, *5*, eaau9490.
[148] H. Cui, R. Hensleigh, D. Yao, D. Maurya, P. Kumar, M. G. Kang, S. Priya, X. R. Zheng, *Nature materials* **2019**, *18*, 234.
[149] Y. Kim, H. Yuk, R. Zhao, S. A. Chester, X. Zhao, *Nature* **2018**, *558*, 274.
[150] H. Wei, Q. Zhang, Y. Yao, L. Liu, Y. Liu, J. Leng, *ACS applied materials & interfaces* **2017**, *9*, 876.
[151] H. Shinoda, S. Azukizawa, K. Maeda, F. Tsumori, *Journal of The Electrochemical Society* **2019**, *166*, B3235.
[152] H. Wei, X. Cauchy, I. O. Navas, Y. Abderrafai, K. Chizari, U. Sundararaj, Y. Liu, J. Leng, D. Therriault, *ACS applied materials & interfaces* **2019**, *11*, 24523.
[153] L.-H. Shao, B. Zhao, Q. Zhang, Y. Xing, K. Zhang, *Extreme Mechanics Letters* **2020**, 100793.

[154] M. Farhang, Seyed, L. Xun, N. Jun, *Materials & Design* **2017**, *122*, 42.
[155] Z. Zhang, K. G. Demir, G. X. Gu, *International Journal of Smart and Nano Materials* **2019**, *10*, 205.
[156] T. Ahlfeld, G. Cidonio, D. Kilian, S. Duin, A. R. Akkineni, J. I. Dawson, S. Yang, A. Lode, R. O. C. Oreffo, M. Gelinsky, *Biofabrication* **2017**, *9*, 034103.
[157] T. Fujita, *Science and technology of advanced materials* **2017**, *18*, 724.
[158] C. W. Hull, *United States Patent, Appl., No. 638905, Filed* **1984**.
[159] sculpteo, BASF, 2020.
[160] Nanodimension, 2020.
[161] Sciperio, 2020.
[162] K. Associates.
[163] Essentium, 2020.
[164] Fortify, 2020.
[165] E. Micro.
[166] Press, 2020.
[167] A. B. W. P. Affairs, Wright-Patterson AFB, 2019.
[168] U. S. A. Force, (Ed: USAF), 2019.
[169] S. Cao, Observer, 2020.
[170] U. Fasel, D. Keidel, L. Baumann, G. Cavolina, M. Eichenhofer, P. Ermanni, *Manufacturing Letters* **2020**, *23*, 85.
[171] Y. W. D. Tay, B. Panda, S. C. Paul, N. A. Noor Mohamed, M. J. Tan, K. F. Leong, *Virtual and Physical Prototyping* **2017**, *12*, 261.
[172] S. Siebert, J. Teizer, *Automation in Construction* **2014**, *41*, 1.
[173] T. D. Ngo, A. Kashani, G. Imbalzano, K. T. Q. Nguyen, D. Hui, *Composites Part B: Engineering* **2018**, *143*, 172.